\documentclass[letter]{article}

\usepackage{amsthm}
\usepackage{times}
\usepackage{url}
\usepackage{epsfig}
\usepackage{amsmath}
\usepackage{longtable}
\usepackage{amsfonts}
\usepackage{amssymb}
\usepackage{graphicx}
\usepackage{epstopdf}
\usepackage{multirow}
\usepackage{subcaption}
\usepackage{verbatim}
\usepackage{lineno}
\usepackage{enumitem}
\usepackage{color}
\usepackage{hyperref}
\usepackage{xspace}

\usepackage[linesnumbered,ruled]{algorithm2e}

\usepackage{self}

\newtheorem{example}{Example}

\usepackage{hyperref}
\usepackage{xcolor}
\hypersetup{
	colorlinks,
	linkcolor={red!50!black},
	citecolor={blue!50!black},
	urlcolor={blue!80!black}
}

\usepackage[numbers]{natbib}

\graphicspath{{small_img/}}

\begin{document}

\title{Estimating multi-year $24/7$ origin-destination demand using high-granular multi-source traffic data}

\author{Wei Ma, Zhen (Sean) Qian\\
	Department of Civil and Environmental Engineering\\ Carnegie Mellon University, Pittsburgh, PA 15213\\
	\{weima, seanqian\}@cmu.edu}
\maketitle

\begin{abstract}

Dynamic origin-destination (OD) demand is central to transportation system modeling and analysis. The dynamic OD demand estimation problem (DODE) has been studied for decades, most of which solve the DODE problem on a typical day or several typical hours. There is a lack of methods that estimate high-resolution dynamic OD demand for a sequence of many consecutive days over several years (referred to as 24/7 OD in this research). Having multi-year 24/7 OD demand would allow a better understanding of characteristics of dynamic OD demands and their evolution/trends over the past few years, a critical input for modeling transportation system evolution and reliability. This paper presents a data-driven framework that estimates day-to-day dynamic OD using high-granular traffic counts and speed data collected over many years. The proposed framework statistically clusters daily traffic data into typical traffic patterns using t-Distributed Stochastic Neighbor Embedding (t-SNE) and k-means methods. A GPU-based stochastic projected gradient descent method is proposed to efficiently solve the multi-year 24/7 DODE problem. It is demonstrated that the new method efficiently estimates the $5$-minute dynamic OD demand for every single day from $2014$ to $2016$ on I-5 and SR-99 in the Sacramento region. The resultant multi-year 24/7 dynamic OD demand reveals the daily, weekly, monthly, seasonal and yearly change in travel demand in a region, implying intriguing demand characteristics over the years.

\end{abstract}

\section{Introduction}

The increasing complexity and inter-connectivity of mobility systems call for  large-scale deployment of dynamic network models that encapsulate traffic flow evolution for system-wide decision making. As an indispensable component of dynamic network models, time-dependent Origin-Destination (OD) demand plays a key role in transportation planning and management. Obtaining accurate and high-resolution time-dependent OD demand is notoriously difficult, though the dynamic OD estimation (DODE) problem has been intensively studied for decades. A number of DODE methods have been proposed, most of which aim at estimating dynamic OD demand for a typical day or even several hours on a typical day. To our best knowledge, there is a lack of research estimating dynamic OD demand for a long time period over the years. The OD demand and its behavior, though are generally repetitive in an aggregated view, can vary from day to day. The day-to-day variation of OD demand would need to be considered in estimate OD demand for a long period of many consecutive days. For example, estimating the dynamic OD demand for every $5$-minutes in an entire year is computationally implausible using most of the existing DODE methods. In view of this, this paper presents an efficient data-driven approach to estimate time-dependent OD demand using high-granular traffic flow counts and traffic speed data collected over many years.

Dynamic OD demand represents the number of travelers departing from an origin at a particular time interval heading for a destination. It reveals traffic demand level, and is critical input for estimating and predicting network level congestion in a region. In addition, policymakers can understand the travelers' departure patterns and daily routines through the day-to-day OD demand. As a result, many Advanced Traveler Information Systems/Advanced Traffic Management Systems (ATIS/ATMS) require accurate time-dependent OD demand as an input. A tremendous number of studies estimate time-dependent OD demand using observed traffic data which includes traffic counts, probe vehicle data and Bluetooth data. Oftentimes those data collected over multiple days are taken daily average before being input to dynamic network models, which represent the average traffic pattern and OD demand on a typical day.

With the development of cutting edge sensing technologies, many traffic data can be collected in high spatial and temporal granularity at a low cost. For example, traffic count and traffic speed for a road segment of 0.1 mile can be sensed and updated every $5$ minutes throughout the year. This is a $12 \times 24 = 288$ dimension of counts/speed data for a single road segment on one day. Most of existing DODE methods become computationally inefficient or even implausible when dealing with large-scale networks with thousands of observed road segments and thousands of days of high dimensional data. How to efficiently obtain high-resolution OD demand on a daily basis over many years remains technically challenging. In this research, we estimate high-resolution dynamic OD demand for a sequence of many consecutive days over several years, referred to 24/7 OD demand throughout this paper.

Dynamic OD estimation (DODE) was formulated as either a least square problem or a state-space model. \citet{cascetta1993dynamic} extended the concepts of static OD estimation problem and formulated a generalized least square (GLS) based framework for estimating dynamic OD demands. \citet{tavana2001internally} proposed a bi-level optimization framework which solves for a GLS problem in the upper level with a dynamic traffic assignment (DTA) problem in the lower level. The bi-level formulations for OD estimation problem were also discussed by \citet{nguyen1977estimating, leblanc1982selection, fisk1989trip, yang1992estimation, florian1995coordinate, jha2004development} for static OD demand. \citet{zhou2003dynamic} extended the bi-level formulation to incorporate multi-day traffic data. To implement efficient estimation algorithms on real-time traffic management systems, \citet{bierlaire2004efficient} proposed a least square based real-time OD estimation/prediction framework for large-scale networks. \citet{zhou2007structural, ashok2000alternative} established a state-space model for real-time OD estimation based on on-line traffic data feeds. \citet{hazelton2008statistical} built a statistical inference framework using Markov chain Monte Carlo algorithm for generating posterior OD demand.

The bi-level OD estimation framework can be solved using heuristically computed gradient, convex approximation or gradient free algorithms. \citet{yang1995heuristic} proposed two heuristic approaches for the bi-level OD estimation problem, the iterative estimation-assignment (IEA) algorithms and sensibility-analysis based algorithm (SAB). \citet{josefsson2007sensitivity} further improved the sensitivity analysis procedures adopted in SAB process. A Dynamic Traffic Assignment (DTA) simulator is also used to determine the numerical derivatives of link flows. \citet{balakrishna2008time, cipriani2011gradient} fitted such an estimation process into a stochastic perturbation simultaneous approximation (SPSA) framework. \citet{lee2009new, vaze2009calibration, ben2012dynamic, lu2015enhanced, tympakianaki2015c, antoniou2015w} further enhanced the SPSA based methods. \citet{verbas2011time} compared different gradient based methods to solve the bi-level formulation of DODE problem. \citet{flotterod2011bayesian} proposed a Bayesian framework that calibrates the dynamic OD using agent-based simulators. In addition to numerical solutions, research has been looking into computing the analytical derivatives for the lower-level formulations \citep{ghali1995model,frederix2011new, qian2012system,qian2011computing}. Other machine learning and computational technologies are also employed to enhance the efficiency of OD estimation methods \citep{kim2001origin, kattan2006noniterative, huang2012computational, xu2014mesoscopic}.

The general bi-level formulation for OD estimation is proved to be non-continuous and non-convex, and thus its scalability is limited. \citet{nie2008variational, nie2010relaxation} formulated a single-level static and dynamic OD estimation framework that incorporates User Equilibrium (UE) path flows solved by the variational inequality, which is further improved by \citet{shen2012new} under the static cases. Recently, \citet{lu2013dynamic} formulated a Lagrangian relaxation-based single-level non-linear optimization to estimate dynamic OD demand.

A large number of data sources are feeding to DODE methods. \citet{zhang2008estimating} evaluated the roles of count data, speed data and history OD data in the effectiveness of DODE. \citet{van1997dynamic, antoniou2004incorporating, zhou2006dynamic, rao2018origin} used automated vehicle identification (AVI) data together with flow counts to estimate dynamic OD demand. Emerging technologies such as Bluetooth \citep{barcelo2010travel}, mobile phone location \citep{calabrese2011estimating, iqbal2014development}, probe vehicles \citep{antoniou2006dynamic} data were also employed to estimate dynamic OD demands.

Two important issues are yet to be addressed. Firstly, many existing DODE methods \citep{ashok2000alternative, josefsson2007sensitivity, nie2008variational, lu2013dynamic, lu2015enhanced} require a dynamic traffic loading (DNL) process (either microscopic or mesoscopic) to endogenously encapsulate the traffic flow evolution and congestion spillover. As the DNL process requires relatively high computational budget, it can take hours to estimate dynamic OD demand on a network of thousands of links/nodes for a single day. Not only does it have hard time converging under the data fitting optimization problem, but estimating the $24/7$ OD demand for several years becomes computationally impractical. The other issue is that most studies estimate OD demand for a few hours or a single day. OD demand varies from day to day, but is also repetitive to some extent. The day-to-day features of OD demand has not be taken into consideration of the DODE methods. For this reason, demand patterns that evolve daily, weekly, monthly, seasonally and yearly have not been explored, despite of high-granular data collected over many years.

In this paper, we develop a data-driven framework that estimates multi-year 24/7 dynamic OD demand using traffic counts and speed data collected over the years. The framework builds the relationship between dynamic OD demand and traffic observations using link/path indices matrix, dynamic assignment ratio (DAR) matrix, and route choice matrix. These three matrices enable the estimate framework to circumvent the bi-level formulation, since each of the matrices can be directly calibrated using high-granular real-world data rather than from complex simulation. The proposed framework utilizes data-driven approaches to explore the daily, weekly, monthly and yearly traffic patterns, and group traffic data into different patterns. The proposed estimation framework is computational efficient: 5-min dynamic OD demand for three years can be estimated within hours on an inexpensive personal computer.

In order to address computation issues, this paper uses a Graphics Processing Unit (GPU) which is currently attracting tremendous research interests from various fields. Neural network models can be performed more deeply and widely \citep{szegedy2015going} with GPU computing. It is also widely used in probabilistic modeling \citep{srivastava2012multimodal} and finite element methods \citep{lu2014coarse}. To our best knowledge, this paper is among the first to design and implement GPU computing in the DODE method, since the traditional DODE methods are not suitable for GPU computing. We present a stochastic gradient projection method that well suits the GPU computing framework. As we will show in the case study, the proposed GPU friendly method is over $10$ times more efficient than the state-of-art CPU based method. The implies that GPU computing makes possible to make full use of the massive traffic data comparing to traditional models.

The main contributions of this paper are summarized as follows:
\begin{enumerate}[label=\arabic*)]
	\item It proposes a framework for estimating multi-year 24/7 dynamic OD demand using high-granular traffic flow counts and speed data. It takes into account day-to-day features of flow patterns by defining and calibrating the dynamic assignment ratio (DAR) matrix using real-world data, which enables realistic representation and efficient computing of network traffic flow.
	\item It adopts t-SNE and k-means methods to cluster daily traffic data collected over many years into several typical traffic patterns. The clustering helps better understand typical daily demand patterns and improve the DODE accuracy.
	\item It proposes a stochastic projected gradient descent method to solve the DODE problem. The proposed method is suitable for GPU computation, which enables efficiently estimating high-dimensional OD over many years.
	\item A numerical experiment on a large-scale network with real-world data is conducted. $5$-minute dynamic OD demands for every day from $2014$ to $2016$ are efficiently estimated. As a result, OD demand evolution over the years can be presented and analyzed.
\end{enumerate}

The remainder of this paper is organized as follows. Section \ref{sec:model} discusses the formulation. Section \ref{sec:imple} presents the solution algorithm for the proposed framework. Section \ref{sec:sol} proposes the entire DODE framework. In section~\ref{sec:case}, a real-world experiment for estimating $5$-minute dynamic OD from $2014$ to $2016$ on a regional Sacramento Network is presented. Finally, conclusions are drawn in Section \ref{sec:con}.

\section{The model}
\label{sec:model}

In this section, we present a framework that utilizes the high-granular traffic counts and speed data to estimate $24/7$ dynamic OD. We first model and discretize continuous-time traffic flow evolution on general networks. The dynamic assignment ratio (DAR) matrix is proposed to characterize the traffic flow evolution in discrete time. Unsupervised dimension reduction and clustering methods are adopted to group data of multiple years into several typical traffic patterns. We use the Logit-based route choice model to characterize travelers' behavior in each cluster. Finally, we formulate the DODE as a high-dimensional non-negative least square (NNLS) problem and propose an efficient solution algorithm.

\subsection{Notations}
Please refer to Table~\ref{tab:notation}. The hat symbol, $\hat{\cdot}$, indicates the variable is an estimator for the true (unknown) variable.

\begin{longtable}{p{3cm}p{10cm}}
	\caption{\footnotesize List of notations}
	\label{tab:notation}
	\endfirsthead
	\endhead	
	$A$ & The set of all links\\
	$A^o$ & The set of links with flow observations\\
	$K_q$ & The set of all OD pairs\\
	$K_{rs}$ & The set of all paths between OD pair $rs$\\
	$\delta_{rs}^{ka}$ & Path/link incidence for $k$th path in OD pair $rs$ and link $a$   \\

\multicolumn{2}{c}{}\\
\multicolumn{2}{c}{\textbf{Variables in continuous time }}\\
	$t_1$ & The departure time of path flow or OD flow\\
	$t_2$ & The arrival time at the tail of link\\
    $T_1$ & The set of all possible departure time from any path and link \\
    $T_2$ & The set of all possible arrival time at all links\\
	$f_{rs}^k(t_1)$ & The $k$th path flow rate for OD pair $rs$ at time $t_1$ \\
	$x_a(t_2)$ & The flow rate at the tail of link $a$ at time $t_2$\\
	$q_{rs}(t_1)$ & The flow rate of OD pair $rs$ at time $t_1$\\
	$c_{rs}^k(t_1)$ & The path cost for path $k$ for OD pair $rs$ departing at time $t_1$\\
	$p_{rs}^k(t_1)$ & The portion of choosing path $k$ in all paths between OD pair $rs$ at time $t_1$\\
\multicolumn{2}{c}{}\\
\multicolumn{2}{c}{\textbf{Variables in discrete time }}\\
	$h_1$ & The index of departure time interval of path flow or OD flow\\
	$h_2$ & The index of arrival time interval at the tail of link\\
	$\bar{f}_{rs}^{kh_1}$ & The $k$th path flow rate for OD pair $rs$ in time interval $h_1$ \\
	$\bar{x}_a^{h_2}$ & The flow rate at the tail of link $a$ in time interval $h_2$\\
	$\bar{q}_{rs}^{h_1}$ & The flow rate of OD pair $rs$ in time interval $h_1$\\
	$\bar{p}_{rs}^{kh_1}$ & The portion of choosing path $k$ in all paths between OD pair $rs$ in time interval $h_1$\\
	$\rho_{rs}^{ka}(h_1, h_2)$ & The portion of the $k$th path flow departing within time interval $h_1$ between OD pair $rs$ which arrives at link $a$ within time interval $h_2$ (namely, an entry of the DAR matrix)

	

\end{longtable}

\subsection{Model the continuous time traffic flow}

\label{sec:flow}
Before proposing the estimation method, we first formulate the model for continuous time traffic flow  on general networks. We denote the path flow $f_{rs}^{k}(t_1)$ as the $k$th path flow rate for OD pair $rs$ at time $t_1$ and link flow $x_a(t_2)$ as the flow rate at the tail of link $a$ at time $t_2$. The relationship between path flow and link flow is presented by Equation~\ref{eq:linkpath}.

\begin{eqnarray}
\label{eq:linkpath}
x_a(t_2) &=& \int_{t_1 \in T_1} \left( \sum_{rs \in K_q} \sum_{k \in K_{rs}} \delta_{rs}^{ka}(t_1, t_2) f_{rs}^k(t_1) \right)dt_1\nonumber\\
&=&\sum_{rs \in K_q} \sum_{k \in K_{rs}} \int_{t_1 \in T_1}   \delta_{rs}^{ka}(t_1, t_2) f_{rs}^k(t_1) dt_1
\end{eqnarray}
where $K_q$ is the set of all OD pairs, and $K_{rs}$ is the path set for OD pair $rs$. $T_1$ is the set of possible departure time for any path and link. In this paper we always denote departure time of path flow or OD flow as $t_1$, and the arrival time at the tail of link as $t_2$, respectively. The time-dependent path/link indices matrix $\delta_{rs}^{ka}(t_1, t_2)$ is defined as follows:

\begin{eqnarray}
\delta_{rs}^{ka}(t_1, t_2) = \begin{cases}
1 & \text{if path flow $f_{rs}^k(t_1)$ arrives at the tail of link $a$ at time $t_2$}\\
0 & \text{else}
\end{cases}
\end{eqnarray}

Assuming the traffic flow is FIFO (First-In-First-Out) and continuous, the arrival time of all departure flows can be determined explicitly. Therefore, the time-dependent path/link indices matrix can be simplified as in Equation~\ref{eq:index}.
\begin{eqnarray}
\label{eq:index}
\delta_{rs}^{ka}(t_1, t_2) = \begin{cases}
\delta_{rs}^{ka} & \text{if $t_1 = \tau_{rs}^{ka}(t_2)$} \\
0 & \text{else}
\end{cases}
\end{eqnarray}
where $\delta_{rs}^{ka}$ is $1$ if path $k$ for OD pair $rs$ passes link $a$ and $0$ otherwise. $\tau_{rs}^{ka}(\cdot)$ is the departure time function for $k$th path in OD $rs$, and $\tau_{rs}^{ka}(t_2)$ is the departure time of $k$th path in OD pair $rs$ arriving at the tail of link $a$ at $t_2$, $\tau_{rs}^{ka}(t_2) \in T_1$. Combining Equation~\ref{eq:linkpath} and Equation~\ref{eq:index} by replacing the time-dependent path/link indices matrix with a static path/link indices matrix, the relationship between link flow and path flow can be formulated as Equation~\ref{eq:linkpath2}.
\begin{eqnarray}
\label{eq:linkpath2}
x_a(t_2)
&=&  \sum_{rs \in K_q} \sum_{k \in K_{rs}} \delta_{rs}^{ka} f_{rs}^k\left(\tau_{rs}^{ka}(t_2) \right)
\end{eqnarray}

\begin{figure}[h]
	\centering
	\includegraphics[scale = 1]{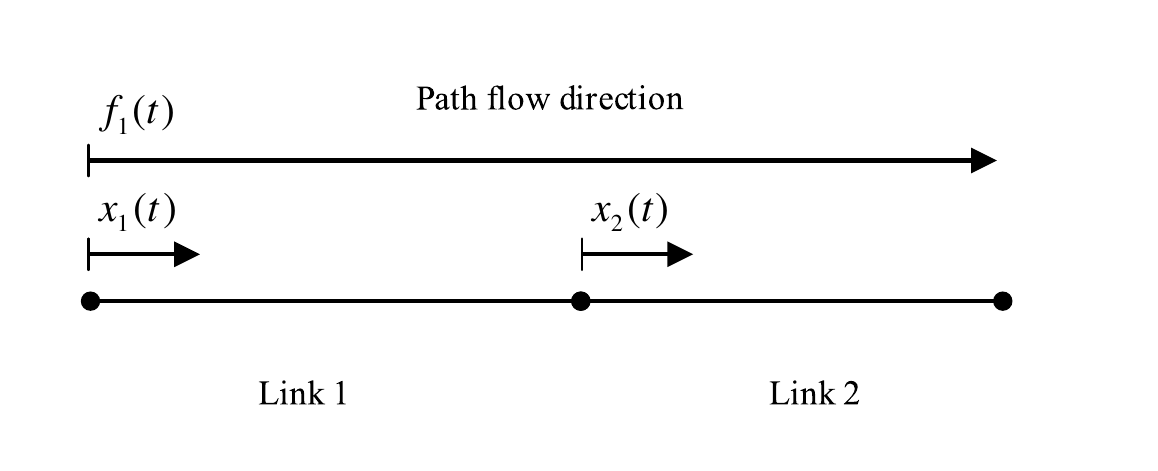}
	\caption{Example of link flow and path flow}
	\label{fig:flow}
\end{figure}

\begin{example}[Link flow and path flow]
Consider a two-link network presented in Figure~\ref{fig:flow}. The path flow is $f_1(t)$, and the link flow for link $1$ and $2$ are $x_1(t)$ and $x_2(t)$, respectively. The travel time to traverse link $1$ is constantly $\Delta t$. Then at the starting time $t_0$, we have
\begin{eqnarray}
x_1(t_0) &=& f_1(t_0)\\
x_2(t_0) &=& 0
\end{eqnarray}
After $\Delta t$, we have
\begin{eqnarray}
x_1(t_0 + \Delta t) &=& f_1(t_0+ \Delta t)\\
x_2(t_0 + \Delta t) &=& f_1(t_0)
\end{eqnarray}
\end{example}

\subsection{Objective function in discrete time}
\label{sec:dar}
The objective function of DODE problem computes the $\ell^2$ norm between the observed link flow $x_a(t_2)$ and the estimated link flow $\hat{x}_a(t_2)$. The estimated link flow is aggregated by the estimated path flows $\hat{f}_{rs}^k(t_1)$, then the optimization problem is presented in Equation~\ref{eq:base}.


\begin{equation}
\label{eq:base}
\begin{array}{rrclcl}
\vspace{5pt}
\displaystyle \min_{\{\hat{f}_{rs}^k(\cdot)\}_{r,s,k}} & \multicolumn{3}{l}{\displaystyle \sum_{a \in A } \int_{t_2 \in T_2} \norm{x_a(t_2) - \hat{x}_a (t_2) }_2^2 dt_2}\\
\textrm{s.t.} & \hat{f}_{rs}^k(t_1) & \geq & 0 & \forall t_1 \in T_1,\forall rs \in K_q, \forall k \in K_{rs}
\end{array}
\end{equation}
where $T_2$ is the set of possible arrival time for all links, which is usually the observation time period for all links. Equation~\ref{eq:base} formulates the objective function on the link set $A$, we can use the observed link set $A^o$ to replac $A$ if only a subset of links are observed. Based on Equation~\ref{eq:linkpath2}, we rewrite the objective function as Equation~\ref{eq:ob}.
\begin{eqnarray}
\label{eq:ob}
L(x, \hat{x})
&=& \sum_{a \in A} \int_{t_2 \in T_2} \norm{ x_a(t_2) - \sum_{rs \in K_q} \sum_{k \in K_{rs}} \delta_{rs}^{ka} \hat{f}_{rs}^k\left(\tau_{rs}^{ka}(t_2) \right) }_2^2 dt_2
\end{eqnarray}

Typically, the data collected from traffic sensors are discretized in terms of time intervals. Therefore, the objective function needs to be discretized as well. We divide the entire time period $T_1 \cup T_2$ into $N$ time intervals, and the sequence of time intervals is denoted as $\{H_h\}_{h=1}^{N}$. We further denote $t^h =  \sup_{t'} \{t' |t' \leq t, \forall t \in H_h\}$, which represents the beginning of each time interval.

\begin{figure}[h]
	\centering
	\includegraphics[scale = 0.8]{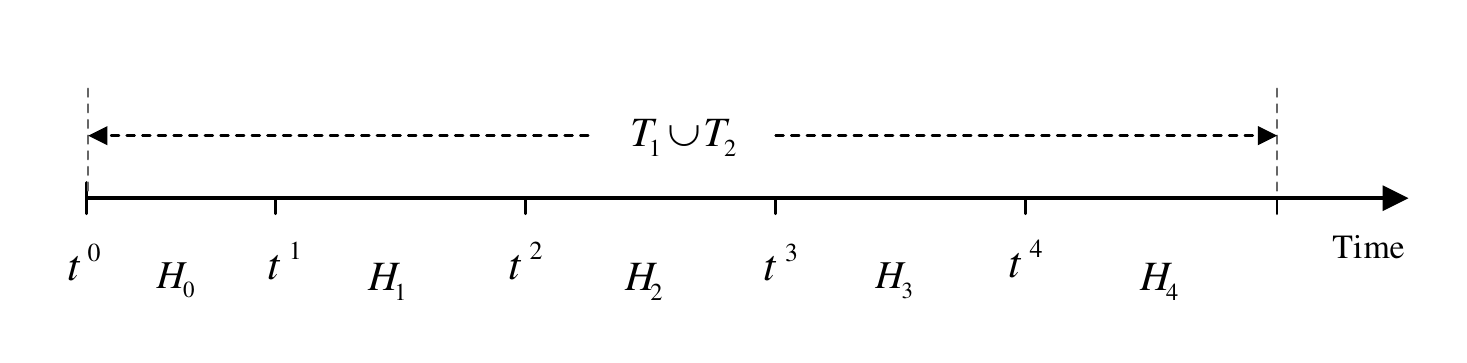}
	\caption{Example of time interval discretization}
	\label{fig:interval}
\end{figure}

\begin{example}[Time interval discretization]
In Figure~\ref{fig:interval}, we discretize the whole time period into $4$ intervals. $H_1, H_2, H_3, H_4$ are the time intervals and $t^1, t^2, t^3, t^4$ are time points denoting the starting time of each time interval.
\end{example}

The discretized objective function is presented in Equation~\ref{eq:obj}.

\begin{eqnarray}
\label{eq:obj}
L(x, \hat{x}) &=& \sum_{a \in A} \int_{t_2 \in T_2} \norm{ x_a(t_2) - \sum_{rs \in K_q} \sum_{k \in K_{rs}} \delta_{rs}^{ka} \hat{f}_{rs}^k\left(\tau_{rs}^k(t_2) \right) }_2^2 dt_2\nonumber\\
\label{eq:objapp}
& \stackrel{Large~N}{\simeq}& \sum_{a \in A}  \sum_{h_2=1}^N \left(  \norm{ \int_{t_2 \in H_{h_2}} x_a(t_2) dt_2 - \sum_{rs \in K_q} \sum_{k \in K_{rs}} \delta_{rs}^{ka} \int_{t_2 \in H_{h_2}} \hat{f}_{rs}^k\left(\tau_{rs}^{ka}(t_2) \right) dt_2 }_2^2  \right)\nonumber\\
&=& \sum_{a \in A}  \sum_{h_2=1}^N \left(  \norm{ \bar{x}_a^{h_2} - \sum_{rs \in K_q} \sum_{k \in K_{rs}} \delta_{rs}^{ka} \sum_{h_1 = 1}^N \left(  \int_{{t_1} \in H_{h_1} \cap \tau_{rs}^{ka}(H_{h_2}) }\hat{f}_{rs}^k(t_1) dt_1\right) }_2^2 \right)\nonumber\\
&=& \sum_{a \in A}  \sum_{h_2=1}^N \left(  \norm{ \bar{x}_a^{h_2} - \sum_{rs \in K_q} \sum_{k \in K_{rs}} \delta_{rs}^{ka} \sum_{h_1 = 1}^N \left(  \rho_{rs}^{ka}  \left(h_1, h_2\right) \hat{\bar{f}}_{rs}^{k h_1 } \right) }_2^2 \right)
\end{eqnarray}
where
\begin{eqnarray}
\bar{x}_a^{h_2}  &=& \int_{t_2 \in H_{h_2}} x_a(t_2) dt_2\\
\hat{\bar{f}}_{rs}^{kh_1} &=& \int_{t_1 \in H_{h_1}} \hat{f}_{rs}^k\left(t_1 \right) dt_1
\end{eqnarray}

We denote $\tau_{rs}^{ka}(H_{h_2})$ as the range of function $\tau_{rs}^{ka}(\cdot)$ with domain being $H_{h_2}$, $\tau_{rs}^{ka}(H_{h_2}) = \{t_1 |t_1 = \tau_{rs}^{ka}(t_2),\forall t_2 \in H_{h_2} \}$. The cumulative link flow $\bar{x}_a^{h_2}$ and cumulative estimated path flow $\hat{\bar{f}}_{rs}^{h_1k}$ are integrated from $x(t_2)$ and $\hat{f}_{rs}^{k}(t_1)$ over time interval $H_{h_1}$ and $H_{h_2}$, respectively. The weight function $\rho_{rs}^{ka}(h_1, h_2)$ denotes the portion of the $k$th path flow departing within time interval $h_1$ between OD pair $rs$ which arrive at link $a$ within time interval $h_2$.
\begin{eqnarray}
\rho_{rs}^{ka}(h_1, h_2) = \frac{\int_{{t_1} \in H_{h_1} \cap \tau_{rs}^{ka}(H_{h_2}) }f_{rs}^k(t_1)dt_1}{\bar{f}_{rs}^{h_1k}}
\end{eqnarray}

We can use this weight function to trace the discretized path flow $\bar{f}^{h_1k}_{rs}$ to link $a$, as presented in Equation~\ref{eq:reverse}.

\begin{eqnarray}
\label{eq:reverse}
\bar{x}_a^{h_2} &=& \sum_{rs \in K_q} \sum_{k \in K_{rs}} \delta_{rs}^{ka} \sum_{h_1 = 1}^{N}  \rho_{rs}^{ka}(h_1, h_2) \bar{f}_{rs}^{kh_1}
\end{eqnarray}

It can be seen that the discretized objective function approaches to the continuous objective function when $N \to \infty$. The weight function $ \rho_{rs}^{ka}$ reflects the link-level flow progression from time interval $h_1$ to $h_2$. The  flow progression and evolution aggregated at the link level can be captured by the time-varying link-level traffic speed and counts. However, its evolution within each link, such as within-link shockwave, can be hardly calibrated or learned unless trajectory level data are available. In fact, link-level flow evolution is proven to be realistic, stable and efficient \citep{jin2012link}. Thus, in this research, we assume vehicles on the network are evenly spread in space and link flow rate at the tail of each link within each time interval is also constant (evenly spread in time), resulting the weight function $ \rho_{rs}^{ka}$ presented in Equation~\ref{eq:unif}.
\begin{eqnarray}
\label{eq:unif}
f_{rs}^k(t_1) = \frac{1}{|H_{h_1}|} \bar{f}_{rs}^{kh_1}, \forall t_1 \in H_{h_1}
\end{eqnarray}

The formulation \ref{eq:unif} is further simpled using equal time intervals, as presented by $\Delta H := |H_h|, \forall h=1, \cdots, n$. Then we are ready to present the dynamic assignment ratio (DAR) as in Equation~\ref{eq:dar}.
\begin{eqnarray}
\label{eq:dar2}
\rho_{rs}^{ka}(h_1, h_2)
&=& \frac{|\tau_{rs}^{ka} (H_{h_2}) \cap H_{h_1} |} {|H_{h_1}|}\\
\label{eq:dar}
&=& \frac{|\left(\tau_{rs}^{ka}\right)^{-1} (H_{h_1}) \cap H_{h_2} |} {|\left(\tau_{rs}^{ka}\right)^{-1} (H_{h_1})|}
\end{eqnarray}
where $\left(\tau_{rs}^{ka}\right)^{-1}(\cdot)$ is the inverse function of $\tau_{rs}^{ka}(\cdot)$ since $\tau_{rs}^{ka}(\cdot)$ is monotonically increasing based on the FIFO rule. $\left(\tau_{rs}^{ka}\right)^{-1} (H_{h_1})$ represents the range of function $\left(\tau_{rs}^{ka}\right)^{-1}$ with domain being $H_{h_1}$.
For each path $f_{rs}^k$, Equation \ref{eq:dar} can be interpreted as the portion of vehicles arriving at link $a$ in time interval $h_2$ among all the vehicles departing at interval $h_1$. As we assumed that the vehicles are spread evenly in time and space, the portion $\rho_{rs}^{ka}(h_1, h_2)$ can be computed either at departing time \ref{eq:dar2} or at arriving time \ref{eq:dar}.
The DAR matrix is computed through the weight function $\rho_{rs}^{ka}(\cdot, \cdot)$.

\begin{figure}[h]
	\centering
	\includegraphics[scale = 0.6]{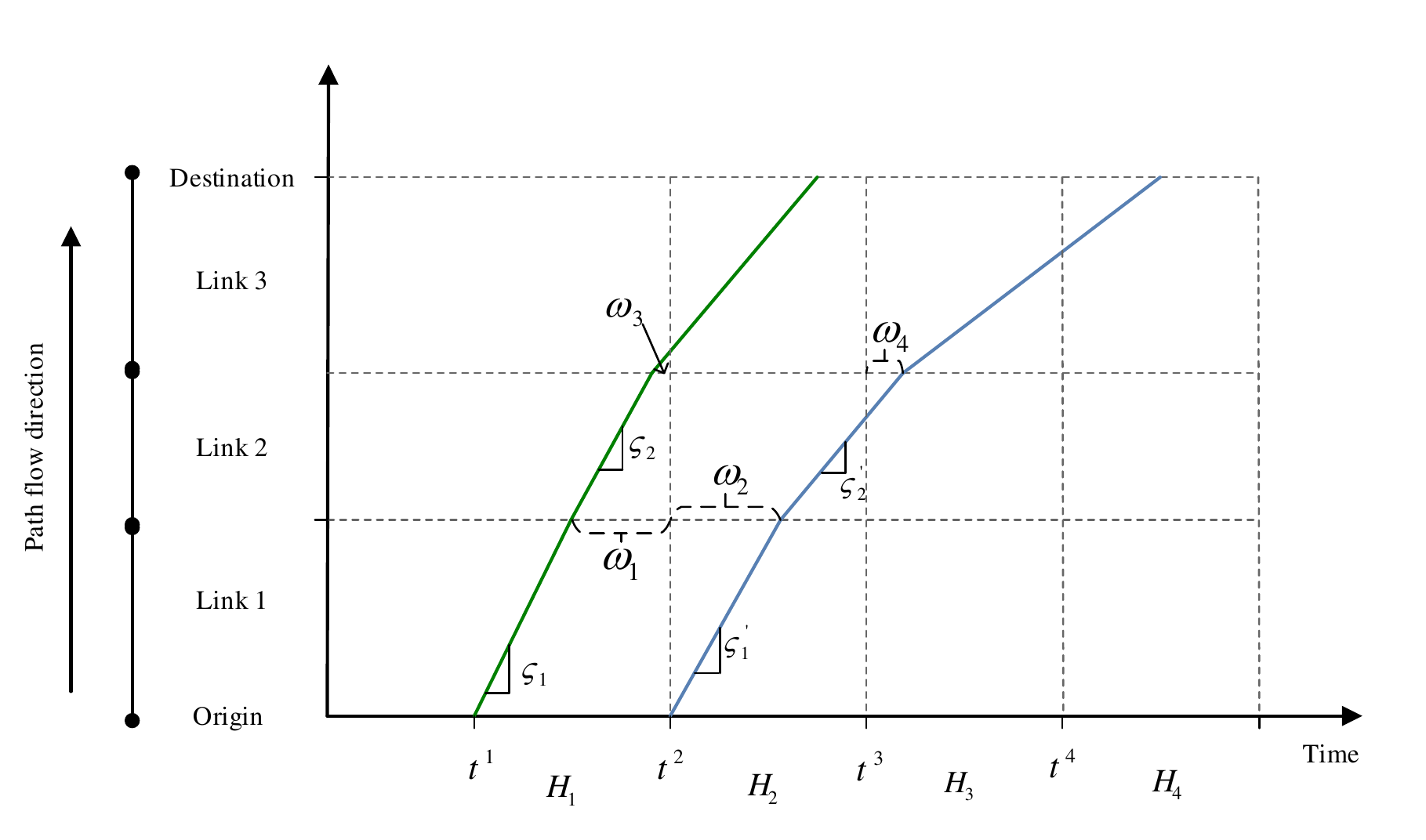}
	\caption{Example of computing the DAR matrix}
	\label{fig:portion}
\end{figure}

\begin{example}[DAR matrix computation]
\label{ex:dar}
As presented in Figure~\ref{fig:portion}, we demonstrate an example for computing the DAR matrix in a three link network. The path flow $f_{rs}^k$ passes three links $x_1, x_2, x_3$ on the network. To compute non-zero entries of the DAR matrix with $h_1 =1$, we derive the trajectories of path flow departing at time $t^1$ and $t^2$. The speeds of links are the slopes of the trajectory, which are denoted as $\zeta_1, \zeta_2, \zeta_1', \zeta_2'$. The probe vehicle speeds of links are available from various sources, such as HERE, INRIX and TomTom. We plot the two approximate trajectories of the leading vehicle departing from the origin at time $t^1$ and $t^2$, and measure the length of each time segment as $\omega_1, \omega_2, \omega_3, \omega_4$.  Based on the definition of $\left(\tau_{rs}^{ka}\right)^{-1}$, we have
\begin{eqnarray}
\left|\left(\tau_{rs}^{k1}\right)^{-1}(H_1)\right| &=& |H_1|\\
\left|\left(\tau_{rs}^{k2}\right)^{-1}(H_1)\right| &=& \omega_1 + \omega_2\\
\left|\left(\tau_{rs}^{k3}\right)^{-1}(H_1)\right| &=& \omega_3 + |H_2| + \omega_4\\
\end{eqnarray}

Then the DARs can be computed as follows based on Equation~\ref{eq:dar}.
\begin{eqnarray}
\rho_{rs}^{k1}(1, 1) &=& 1\\
\rho_{rs}^{k2}(1, 1) &=& \frac{\omega_1}{\omega_1 + \omega_2}\\
\rho_{rs}^{k2}(1, 2) &=& \frac{\omega_2}{\omega_1 + \omega_2}\\
\rho_{rs}^{k3}(1, 1) &=& \frac{\omega_3}{\omega_3 + |H_2|+ \omega_4}\\
\rho_{rs}^{k3}(1, 2) &=& \frac{|H_2|}{\omega_3 + |H_2|+ \omega_4}\\
\rho_{rs}^{k3}(1, 3) &=& \frac{\omega_4}{\omega_3 + |H_2|+ \omega_4}
\end{eqnarray}
\end{example}

Given Equation~\ref{eq:dar}, the discrete time objective function is formulated as Equation~\ref{eq:finalobj}:
\begin{eqnarray}
\label{eq:finalobj}
L(x, \hat{x}) \simeq \sum_{a \in A}  \sum_{h_2=1}^N \left(  \norm{ \bar{x}_a^{h_2} - \sum_{rs \in K_q} \sum_{k \in K_{rs}} \sum_{h_1 =1}^N\delta_{rs}^{ka} \rho_{rs}^{ka}(h_1, h_2) \hat{\bar{f}}_{rs}^{h_1k} }_2^2  \right)
\end{eqnarray}

\subsection{Link/path travel time}

In previous sections, we derive the objective function based on the DAR matrix. As shown in Example~\ref{ex:dar}, the DARs are computed through $\omega_1, \omega_2, \omega_3, \omega_4$. These variables can be computed based on the link travel time, for example
\begin{eqnarray}
\omega_1 = t^2 - \left(t^1 + c_1(t^1) \right)
\end{eqnarray}

In a general form, let $c_a(t)$ denote the travel time of link flow for $a$ departing from the tail of link at time $t$. We denote $c_{rs}^k(t)$ as the travel time of path flow $k$ in OD pair $rs$ departing at time $t$. Let $\alpha_{rs}^{k}$ represent the sequence of links passed by flow $f_{rs}^k$, $\alpha_{rs}^k(a)$ represent the $a$th link in sequence $\alpha_{rs}^{k}$ , and $\beta_{rs}^k$ represents the number of links passed by flow $f_{rs}^k$. Then $c_{rs}^k(t)$ can be calculated by Equation~\ref{eq:cost}.
\begin{eqnarray}
\label{eq:cost}
c_{rs}^k(t_1) = c_{\alpha_{rs}^{k}(\beta_{rs}^k)}\left(c_{\alpha_{rs}^{k} \left(\beta_{rs}^k - 1\right)} \left( \cdots  \left( c_{\alpha_{rs}^{k}(1)}(t_1) \right) \right)\right)
\end{eqnarray}

We note the link travel time can be obtained from either dynamic network loading models (traffic simulation) or the real-world data. In this research, we use the speed data from probe vehicles (such as INRIX or HERE) to circumvent the simulation process. The link/path travel time can be directly calibrated from the high-granular probe vehicle speed data.

\subsection{Traffic pattern clustering}
\label{sec:cluster}
In the following sections, we will build the relationship between dynamic OD flow and dynamic path flow. Behavior models determines the route choice portions based on the traffic conditions and travelers perception errors, which are used to distribute OD flow onto different paths. Travelers' route choices are likely to be stable when traffic conditions are recurrent. In this research, we speculate that there exist several typical repetitive traffic conditions at the network level, each of which carries weekday/weekend, seasonal or other demand/supply characteristics. In each typical traffic pattern, we assume the network condition follows a statistical equilibrium defined by \citet{GESTA,ma2018statistical}. Travelers will select their route based on the traffic pattern they observe historically, and  their route choice portions remains stable for those days with the same typical traffic pattern. To estimate the route choice portions in each traffic pattern, we first cluster the traffic data into patterns using day-to-day traffic data in this section. Then the route choice portions for each pattern are estimated based on a generalized route choice model in the following section.

In addition to statistical equilibrium approach, the day-to-day traffic assignment model can also be used to utilize temporal correlation of traffic patterns, and the OD demand can be estimated by a filtering approach. One novelty that stems from the statistical equilibrium approach, to be further examined in the next step, is that the weekly/monthly/seasonal O-D variation can be learned directly from real-world data rather than being a prior to be imposed to the day-to-day dynamics model. In this paper we focus on the statistical equilibrium approach to modeling the temporal correlation of traffic patterns.

To cluster the traffic patterns, t-SNE (t-Distributed Stochastic Neighbor Embedding) is adopted to project high-dimensional traffic data points to low dimensional feature space. K-means method is then used to cluster the data points in the feature space. Each cluster obtained from k-means method represents traffic patterns under different traffic conditions.

\subsubsection{Dimension reduction and data visualization}
For a traffic state variable, e.g. link flow from all sensors on a network, we adopt state-of-art dimension reduction method t-SNE (t-Distributed Stochastic Neighbor Embedding) to project traffic state variables to low dimensional space. The dimension reduction process can significantly reduce the influence of noise and outliers to the clustering methods. The t-SNE method minimizes Kullback-Leibler divergence $C$ between a joint probability distribution $P$ in the high-dimensional space and a joint probability distribution $Q$ in the low-dimensional space, as presented in Equation~\ref{eq:kl}.

\begin{eqnarray}
\label{eq:kl}
C = KL\left(P ||Q\right) = \sum_i \sum_j \mu_{ij} \log \frac{\mu_{ij}}{\nu_{ij}}
\end{eqnarray}
where $i,j$ are the indices of the data. $\mu_{ij}$ and $\nu_{ij}$ measure the pair-wise similarity between data points, which are defined as:

\begin{eqnarray}
\mu_{ij} &=& \frac{\exp\left(-\norm{\chi_i - \chi_j}^2/ 2\sigma^2\right)}{\sum_{i' \neq j'} \exp\left(-\norm{\chi_{i'} - \chi_{j'}}^2/2\sigma^2\right)}\\
\nu_{ij} &=& \frac{\left(1 + \norm{\psi_i - \psi_j}^2\right)^{-1}}{\sum_{i' \neq j'} \left(1 + \norm{\psi_{i'} - \psi_{j'}}^2\right)^{-1}}
\end{eqnarray}
where $\chi_i$ are data points from original high-dimensional space and $\psi_i$ are data points from low-dimensional space that we want. $\psi_i$ is assumed to follow a Student t-distribution with one degree of freedom as one heavy-tailed distribution in low-dimensional space. The computational and space complexity of t-SNE are $\O(n^2)$, but it can be efficiently solved using stochastic gradient descent (SGD) methods with limited number of iterations.

In this research, t-SNE is used as the dimension reduction method, but other clustering methods, such as principal component analysis (PCA), can be potentially adopted as well for the same purpose \citep{chen2018spatial}. Among all the dimension reduction methods, t-SNE is able to handle the non-linear relationship between variables and hence form smaller groups compared to other methods \citep{garcia2013stability}. Many studies have demonstrated the effectiveness of t-SNE in handling very high-dimensional datasets \citep{booth20163d, th2015evaluating}. in the numerical example, we also compare the t-SNE with other PCA-based methods and demonstrate the effusiveness of t-SNE.

We set  $\chi_i$ as the vector of observed traffic counts or traffic speed on each day and $i$ denotes the index of the dates. $\chi_i$ is a one-dimensional vector with length $N\times O$, where $N$ is the number of time intervals in a day and $O$ is the number of observations per time interval. Then we minimize the objective function $C$ to search for the low dimensional feature $\psi_i$, where $i$ also denotes the index of dates. Then we are able to use the feature $\psi_i$ to represent the high dimension variable $\chi_i$ for each day.

One important feature of the projected dimension by t-SNE is that it has state-of-art  visualization properties of data. The low dimensional space not only retains the local structure of the data, but also reveals the global structure in the high dimensional space.

\subsubsection{Clustering}

Clustering methods group day-to-day traffic data into different patterns. Since t-SNE projects traffic data onto low dimensional feature space, which reflects the structure of high dimensional space. Even a simple clustering method works well on the feature space. In this research, we adopt k-means method to cluster the feature space.

We project traffic speed and traffic counts to feature space and build the clustering models, respectively. Suppose there are data available for $D$ days, we will have $U$ clusters for speed data and $V$ clusters for count data after t-SNE and K-means. Then we define $U\times V$ clusters as $\{(u,v)| u \in U, v \in V\}$.

The intuition behind the clustering process is two-fold: 1) Count data and speed data have different structures in the high dimensional space. Count data have larger variance than the speed data. Thus, parameter tuning for t-SNE should be different for count versus speed data. 2) Travelers' route choice is a combined decision process based on the traffic demand (count data) and traffic congestion (speed data) together. Hence we use the composite of count clusters and speed clusters to represent different patterns.

The clustering method we adopt is data-driven. Hard-coding the clusters using prior knowledge such as weekday/weekends or seasons is not necessary. Later we will show in the case study that the clustering results actually reflect not only weekday/weekend traffic patterns, but also other non-trivial factors such as incidents and events.

\subsection{Route choice portions}
\label{sec:route}
For each traffic pattern, we compute the route choice portions for all OD pairs. Define route choice portion $p_{rs}^k(t_1)$ such that it distributes OD demand $q_{rs}(t_1)$ to path flow $f_{rs}^k(t_1)$ by Equation~\ref{eq:ODpath}.
\begin{eqnarray}
\label{eq:ODpath}
f_{rs}^k(t_1) = p_{rs}^k(t_1) q_{rs}(t_1)
\end{eqnarray}
where $p_{rs}^{k}(t_1)$ represents the route choice portion of $k$th path flow in OD pair $rs$ departing at time $t_1$. The time-dependent route choice portion $p_{rs}^{k}(t)$ can be determined through a generalized route choice model, as presented in Equation~\ref{eq:gen_choice}.
\begin{eqnarray}
\label{eq:gen_choice}
\left(p_{rs}^k(t_1)\right)_i = \Psi_{rs}^k\left( \D(i) ; i\right)
\end{eqnarray}
where $\left(p_{rs}^k(t_1)\right)_i$ denotes the route choice portions for $k$th path  in OD $rs$ at time $t_1$ for pattern $i$. $\D(i)$ represents the traffic conditions (flow, travel time, speed, travel time reliability, etc.) of all those days within the pattern $i$. $\Psi_{rs}^k(\cdot)$ is a generalized route choice model that takes any information within the traffic pattern and compute the route choice portion for travelers in $k$th path in OD $rs$. To simplify the notation, we ignore the pattern index $i$ in the rest of the paper.

For instance, we can use a Logit-based model based on mean travel time for each traffic pattern as shown in Equation~\ref{eq:logit}.

\begin{eqnarray}
\label{eq:logit}
p_{rs}^k(t_1) = \frac{\exp\left(-\theta \tilde{c}_{rs}^k (t_1)\right)}{\sum_{k \in K_{rs}}\exp\left(-\theta \tilde{c}_{rs}^k(t_1)\right)}
\end{eqnarray}
where $\tilde{c}_{rs}^k$ represents the mean travel time of path flow $k$ in OD $rs$ departing at time $t_1$ for all days within the cluster (or pattern). $\theta$ is the dispersion factor in Logit model. To discretize the time, we further assume that the route choice portions stay the same in each time interval, then,


\begin{eqnarray}
\label{eq:rt}
\bar{p}_{rs}^{kh_1} :=  p_{rs}^k(t_1), \forall t_1 \in H_{h_1}
\end{eqnarray}

The discrete time link flow and path flow can be formulated as in Equation~\ref{eq:flow}.
\begin{eqnarray}
\bar{f}_{rs}^{kh_1} &=& \int_{t_1 \in H_{h_1}} f_{rs}^k\left(t_1 \right) dt_1 \nonumber\\
&=&  \int_{t_1 \in H_{h_1}} p_{rs}^k(t_1) q_{rs}\left(t_1 \right) dt_1 \nonumber \\
&=& \bar{p}_{rs}^{kh_1} \int_{t_1 \in H_{h_1}} q_{rs}(t_1)  dt_1 \nonumber\\
&=& \bar{p}_{rs}^{kh_1} \bar{q}_{rs}^{h_1} \label{eq:flow}
\end{eqnarray}

\subsection{Estimate the dynamic OD demand}

Now we are ready to present the formulation for solving the DODE problem. Combining Equations \ref{eq:base}, \ref{eq:finalobj} and \ref{eq:flow}, the DODE formulation is presented in Equation~\ref{eq:opt}.

\begin{equation}
\label{eq:opt}
\begin{array}{rrclclc}
\vspace{5pt}
\displaystyle \min_{\{q_{rs}^{h_1}\}_{r,s,h_1}} & \multicolumn{4}{l}{\displaystyle \sum_{a \in A^o}  \sum_{h_2=1}^N \left(  \norm{ \bar{x}_a^{h_2} - \sum_{rs \in K_q} \sum_{k \in K_{rs}} \sum_{h_1 =1}^N \delta_{rs}^{ka} \rho_{rs}^{ka}(h_1, h_2) p_{rs}^{kh_1}  \bar{q}_{rs}^{h_1} }_2^2  \right)} \\
\textrm{s.t.} & \bar{q}_{rs}^{h_1} & \geq & 0 & \forall rs \in K_q, 1\leq h_1 \leq N
\end{array}
\end{equation}

In the  formulation \ref{eq:opt}, link flows $\bar{x}_a^{h_2}$ are observed from sensors, path/link indices matrix $\delta_{rs}^{ka}$ is from network topology in section~\ref{sec:flow}, DAR matrix can be computed through real-time traffic speed data by section \ref{sec:dar} and route choice matrix $p_{rs}^{kh}$ is determined by the clustering results in section~\ref{sec:cluster} and the route choice model in section~\ref{sec:route}.  We can formulate the multi-day 24/7 DODE problem as one large non-negative least square (NNLS) problem by viewing the $T_1 \cup T_2$ as the entire observation time period (e.g., $3$ years in the case study). However, to ensure computational efficiency, a best practice is to decompose the NNLS problem of multiple years into subproblems for each of those days separately. This does not come without a price, though. The vehicles departing at the end of day $1$ and arriving in the beginning day $2$ are overlooked in this simplified process. This is still acceptable in practice since midnight OD is usually minimal and of less interest in general. One nice feature of solving NNLS on the daily basis is that it convenient to utilize the parallel computational power to estimate the dynamic OD of each day separately. In the reminder of this paper, the optimization problem~\ref{eq:opt} applies for each day separately and we simply ignore the index for days.

In formulation \ref{eq:opt}, the link capacity constraints (the estimated link flow should be less and equal than the maximum flow capacity) are not explicitly enforced, since these constraints are usually satisfied by 1) achieving the minimum of the objective function close to zero; and 2) enforcing proper route choice models. As can be seen in the following case study, this is generally satisfied. In practice, if it is not the case, enforcing the link flow capacity as additional linear constraints to formulation \ref{eq:opt}  is straightforward under an iterative balancing framework \citep{zhang2008development}. 

We denote $\Beta$ as the assignment matrix, the entries of $\Beta$ can be computed as in Equation~\ref{eq:assign}.
\begin{eqnarray}
\label{eq:assign}
\Beta_{rs}^{ka}(h_1, h_2) = \delta_{rs}^{ka} \rho_{rs}^{ka}(h_1, h_2) \bar{p}_{rs}^{kh_1}
\end{eqnarray}

Formulation~\ref{eq:opt} is a non-negative least square (NNLS) problem in terms of $x^{h_2}$ and $\Beta$, which can be solved very efficiently in a low dimensional space \citep{lawson1995solving} using the standard NNLS solver. But the standard method can be very inefficient in a high dimensional space, as it computes the inverse of $\Beta^T \Beta$ during the solving process. The dimension of $\Beta^T\Beta$ is usually in billions for a typical DODE problem that estimates daily dynamic OD. In the following section, we will propose a stochastic projected gradient descent method to solve the high-dimensional NNLS problem and implement it on GPU. The DODE problem on a single day can be solved in seconds using this proposed method.

\section{Solution algorithm}
\label{sec:imple}

In previous section, we formulate the 24/7 DODE problem as a non-negative least square (NNLS) problem, as presented in Equation~\ref{eq:nnls}.

\begin{equation}
\label{eq:nnls}
\begin{array}{rrclclc}
\vspace{5pt}
\displaystyle \min_{\bar{q}} & \multicolumn{4}{l}{\displaystyle \norm{\bar{x} - \Beta \bar{q}}_2^2} \\
\textrm{s.t.} & \bar{q}_{rs}^{h_1} & \geq & 0 & \forall rs \in K_q, 1 \leq h_1 \leq N
\end{array}
\end{equation}
where $\bar{x}$ and $\bar{q}$ are the tensor representations of link flows and the OD flows in all time intervals, respectively. $\Beta$ is the assignment matrix. The construction of the tensor representations will be presented in the following section.

With the increasing granularity of traffic data, the dimensions of tensor $x, q$ and matrix $\Beta$ grow quickly. Thus, we have to work on a high dimensional space for the proposed DODE framework. In this section, we discuss the technical details of each component of the solution algorithm that ensures computationally efficient implementation of the proposed framework.

\subsection{Tensor representation}
To enable tensor manipulation and computation during the DODE framework, all the variables involved need to be vectorized. For sparse matrices in the formulation, we use coordinate format sparse representation of the matrices.

For $N$ intervals, denote total number path is $\Pi = \sum_{rs} |K_{rs}|$, $K=|K_q|$. The vectorized variables are presented in Table~\ref{tab:vec}. Multiplications between sparse matrix and sparse matrix, sparse matrix and dense vector are very efficient, especially on multi-core CPUs or Graphics Processing Units (GPU).

\begin{table}[h!]
	\begin{center}
		\caption{DODE framework variable vectorization}
		\label{tab:vec}
		\begin{tabular}{ccccp{3cm}}
			\hline
			Variable & Notations & Dimension & Type & Description\\
			\hline\hline \rule{0pt}{3ex}
			OD flow & $q_{rs}^h$ & $\mathbb{R}^{N|K|}$ & Dense & $k$th OD flow in time interval $h$ is place at entry $(h-1)|K| + k$\\
			\hline \rule{0pt}{3ex}
			Path flow & $f_{rs}^{kh}$ & $\mathbb{R}^{N\Pi}$ & Dense & $k$th path flow in time interval $h$ is placed at entry $(h-1)\Pi + k$\\
			\hline \rule{0pt}{3ex}
			Link flow & $x_a^h$ & $\mathbb{R}^{N|A|}$ & Dense & $k$th link flow in time interval $h$ is placed at entry $(N-1)|A| + k$\\
			\hline \rule{0pt}{3ex}
			DAR matrix & $\rho_{rs}^{ka}(h_1, h_2)$ & $\mathbb{R}^{N|A| \times N\Pi}$ & Sparse & Dynamic assignment ratio of $k$th path in OD $rs$ in time interval $h_1$ for link $a$ in time interval $h_2$ is placed at entry $[(h_2-1)|A| + a, (h_1-1)\Pi + k]$\\
			\hline \rule{0pt}{3ex}
			Link/path indices matrix & $\delta_{rs}^{ka}$ & $\mathbb{R}^{|A| \times \Pi}$ & Sparse & $\delta_{rs}^{ka}$ is $1$ if path $k$ for OD pair $rs$ passes link $a$\\
			\hline \rule{0pt}{3ex}
			Route choice matrix & $p_{rs}^{kh}$ & $\mathbb{R}^{N\Pi \times N|K|}$ & Sparse & Route choice for path $k$ for OD pair $rs$ in time interval $h$ is placed at entry $[(h-1)|\Pi| + k, (h-1)|K| + rs]$\\
			\hline
		\end{tabular}
	\end{center}
\end{table}


\subsection{Constructing the dynamic assignment ratio (DAR) matrix}

\label{sec:dar2}

The assignment matrix $\Beta$ is the multiplication of Link/path indices matrix, DAR matrix and route choice matrix. As shown in Table~\ref{tab:vec}, the largest matrix among the three matrices is the dynamic assignment ratio (DAR) matrix. DAR matrix is constructed by network topology and speed data, and the construction process turns out to be the most time-consuming part in the DODE framework.

The construction process for DAR matrix requires iterations over all departure/arriving time intervals, paths and links. We find a way to construct DAR matrix by only iterating over departure time intervals and paths. The links and arriving time intervals will be iterated implicitly when we compute the travel time of each path. For specific time interval and path, we iterate over all the links in the path from origin to destination and compute the arrival time of each link. Using the arrival time, we can compute assignment ratio and put it to its corresponding entry in DAR matrix.

We can also use multi-process computing to construct DAR matrix for multiple days simultaneously. The parallel construction framework can significantly reduce the total computation time.

\subsection{Non-negative least square on GPU}

After constructing assignment matrix $\Beta$, the 24/7 DODE problem is simplified to a non-negative least square problem presented in Equation~\ref{eq:nnls}. However, solving such NNLS problem in high-dimensional space is non-trivial. For a general network, the dimension of OD vector is usually above ten thousand, and standard NNLS solver \citep{lawson1995solving} is not able to handle such a high dimensional problem.

We propose a stochastic projected gradient descent method to solve the high dimensional NNLS problem. The process of the solution method is presented in Algorithm~\ref{alg:nnls}.

\begin{algorithm}[H]
	\SetKwInOut{Input}{Input}
	\SetKwInOut{Output}{Output}
	
	\underline{NNLS} $(\Beta, y, b, \eta, E)$\;
	\Input{matrix $\Beta$, output $y$, batch size $b$, learning rate $\eta$, number of epoch $E$}
	\Output{$x$ such that $\Beta x = y, x\geq 0$}
	$(n, d) = \Beta.\text{shape}$\;
	Initialize $x \in \mathbb{R}^n$\;
	\For{$\text{iter}\gets1$ \KwTo $E$}{
		permuted\_sequence =  \texttt{permutate}(range($n$))\;
		chunk\_list = \texttt{make\_chunk}(permuted\_sequence, $b$)\;
		\For{$\text{chunk} \in \text{chunk\_list}$}{
			$\Beta_o = \Beta [\text{chunk}, :]$\;
			$g = \Beta_o^T (\Beta_o x - y)$\;
			$x = \texttt{Adagrad}(x, g, \eta)$\;
			$x = \max(x, 0)$
		}
	}
	\caption{Stochastic Projected Gradient Descent (SPGD) method for NNLS}
	\label{alg:nnls}
\end{algorithm}

In the algorithm, the batch size $b$, learning rate $\eta$ and number of epoch $E$ are parameters for the SPGD method. Larger batch size implies better convergence rate but larger memory consumption; learning rate is dependent on the problem scale and larger learning rate implies better convergence rate; and larger number of epoch implies the better solution for the NNLS but longer computational time. The \texttt{permutate} function permutates the sequence in random order, \texttt{make\_chunk} function divide a sequence to small chunks with same size. \texttt{Adagrad} is a variant of stochastic gradient (SGD) descent method, it outperforms the SGD during the experiments. Adagrad is an adaptive step size for SGD that is often used to optimize neural networks. Details of the Adagrad method can be found in \citet{duchi2011adaptive}.

We implemented the proposed Algorithm~\ref{alg:nnls} in PyTorch, all the matrices multiplication can be  evaluated on GPU. As we will show in later section, the implemented method can solve NNLS with a $10$ thousand dimension in seconds.

\section{Estimation framerwork}
\label{sec:sol}

In this section, we present the proposed DODE pipeline given the network topology, speed data and count data. Path set of each OD pair needs to be generated prior to the estimation framework. For small networks, path enumeration is possible. When the networks are large, we can simply enumerate $K$ shortest paths \citep{yen1971finding, eppstein1998finding} for each OD pair and then search for the solution in the prescribed path set.

Count data and speed data need to be cleaned and imputed (if missing) before the estimation framework. Network topology and OD pairs will be converted to a directed graph with weighted edges. The entire DODE framework is summarized as follows,

\ \\
\begin{tabular}{p{4cm}p{10cm}}
	\textbf{DODE framework}&  \\\hline
	\textit{Step 0} & \textit{Data preparation.} Build directed graph representation for networks, enumerate paths for all OD pairs. Prepare link count data and speed data, attach data points to the edges of graph. \\\hline
	\textit{Step 1} & \textit{Constructing DAR matrix.}  Construct DAR matrix using the graph and speed data by Section~\ref{sec:dar} and \ref{sec:dar2}.\\\hline
	\textit{Step 2} & \textit{Traffic data clustering.} Divide the data into different traffic patterns by clustering the speed data and count data using methods presented in section~\ref{sec:cluster}.\\ \hline
	\textit{Step 3} & \textit{Constructing route choice matrix.}  Construct the route choice matrix for each traffic pattern using methods presented in section~\ref{sec:route}. \\ \hline
	\textit{Step 4} & \textit{Constructing observed link flow.} Construct the count data for each day using the notation presented in Table~\ref{tab:vec}.\\\hline
	\textit{Step 5} & \textit{Stochastic Projected Gradient Descent for NNLS.}  Specify learning rate and batch size based on different problem size, conduct Stochastic Projected Gradient Descent for NNLS presented in Algorithm~\ref{alg:nnls} for each day.\\ \hline
	\textit{Step 6} & \textit{Quality check.} Check the goodness of fit for the estimated dynamic OD demand and output the results.\\ \hline
\end{tabular}

\section{Numerical experiment: a Sacramento Regional Network}
\label{sec:case}

In this section, we conduct a case study on I-5 and Hwy-99 towards Sacramento. 5-min count and speed data for the years of 2014 to 2016 are used to estimate $5$-minute dynamic OD demands over $3$  years. Efficiency of the proposed methods and goodness of fit are evaluated. We visualize the evolution of estimated OD demand in several ways and discuss the benefits of the high-granular traffic data.

All the experiments below are conducted on a desktop with Intel Core i7-6700K CPU @ 4.00GHz $\times$ 8, 2133 MHz 2 $\times$ 16GB RAM, GeForce GTX 1080 Ti/PCIe/SSE2, 500GB SSD.

\subsection{Data acquisition and preprocessing}

We first describe the network, traffic count and speed data used in the case study. The data preprocessing involves the graph construction, data geocoding, data cleaning, data imputation and data interpolation.

\subsubsection{Network}

\begin{figure}[h]
	\centering
	\includegraphics[scale = 0.6]{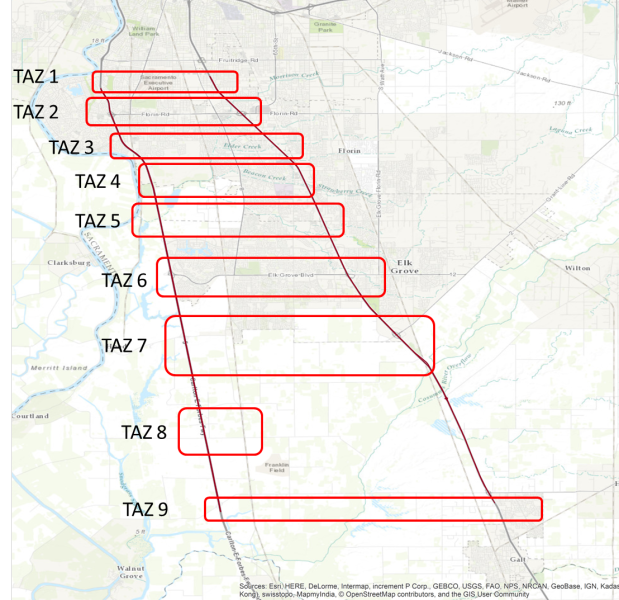}
	\caption{Overview of network and TAZ zones}
	\label{fig:taz}
\end{figure}

I-5 and SR-99 are the two highway corridors in this network. The OD connectors are constructed based on the residence region and interchanges/ramps of two highways. We divide the entire network into $9$ traffic analysis zones (TAZs), and attach one origin and one destination to each TAZ. The overview of all $9$ TAZs are shown in Figure~\ref{fig:taz}.

The $9$ TAZs are across two major highways towards Sacramento downtown. The main purpose of this case study is to characterize the traffic demand in the southern region of Sacramento heading/leaving Sacramento downtown.  Northern regions of TAZ $1$ are not modeled since there are too many highway exits/entrances and local roads, our data are not rich enough to accurately model the demand profile in those regions. The north of TAZ $9$ are not modeled since there is few resident area in this area. We further enumerate all paths to generate the path set for each OD pair.

\subsubsection{Counts}

\begin{figure}[h]
	\centering
	\includegraphics[scale = 0.3]{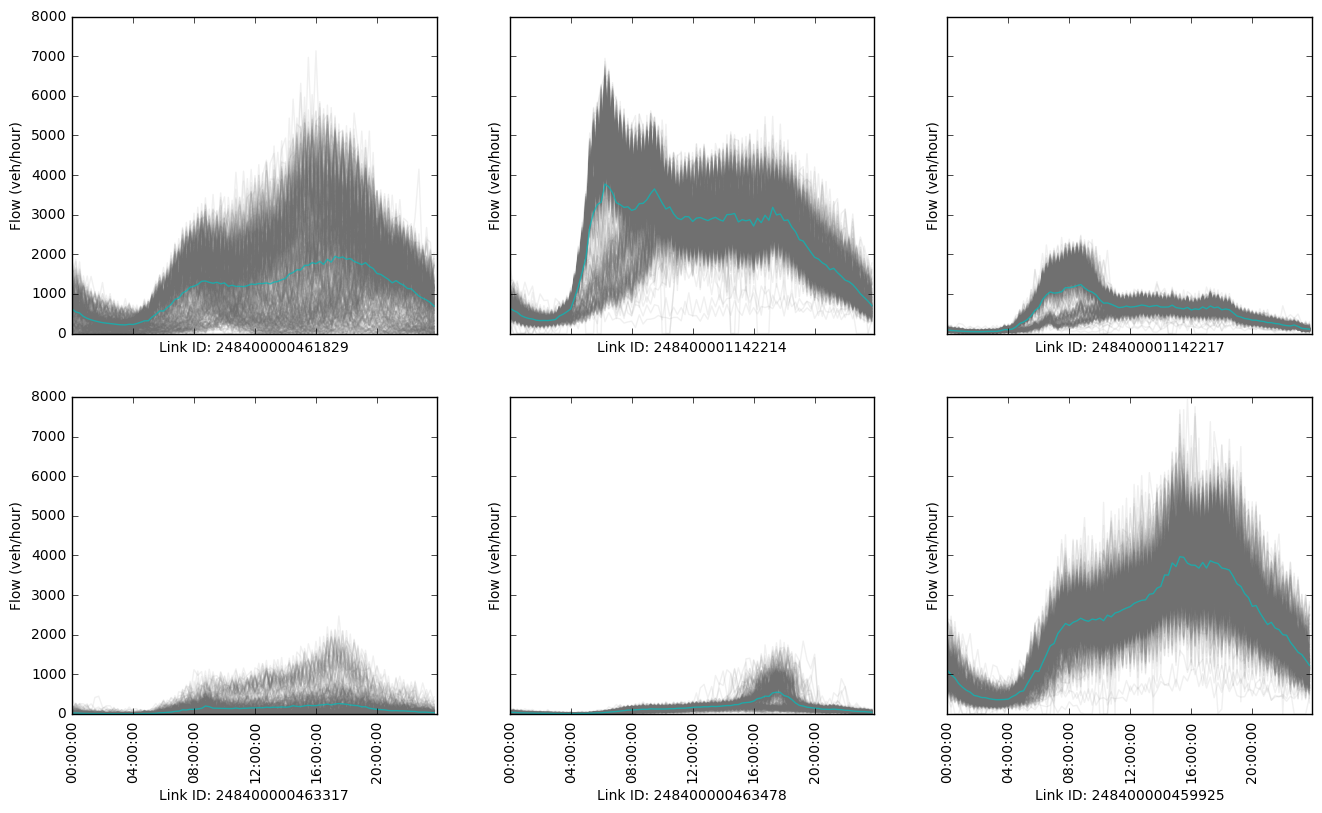}
	\caption{Traffic counts for randomly selected  $6$ sensors}
	\label{fig:count}
\end{figure}

The flow count raw data are obtained from Caltrans Performance Measurement System (PeMS), which is a combined source from various types of vehicle detector stations, including inductive loops, side-fire radar, and magnetometers. The count data contain the traffic counts from $94$ locations in every $5$ minutes for $3$ years. There exist several sensors on the same road segment. In this case, we take the average of counts for that segment. On each day, there are $60min / 5min \times 24 hour = 288$ time intervals, thus the traffic count data for each day is a vector in $\mathbb{R}^{288}$. We randomly select $6$ locations and visualize the day-to-day traffic counts. The average traffic counts over the $3$ years for each time interval are also plotted in Figure~\ref{fig:count}. Each grey time-of-day trace represents traffic counts over one day, and the blue line represents the average daily time-of-day traffic counts over three years.

As can be seen from Figure~\ref{fig:count}, traffic counts data on most of days follow similar trends but contain large day-to-day variation. Some sensors pick up morning peaks and afternoon peaks, while others can only capture either or neither of the traffic peaks.

\subsubsection{Speeds}

\begin{figure}[h]
	\centering
	\includegraphics[scale = 0.65]{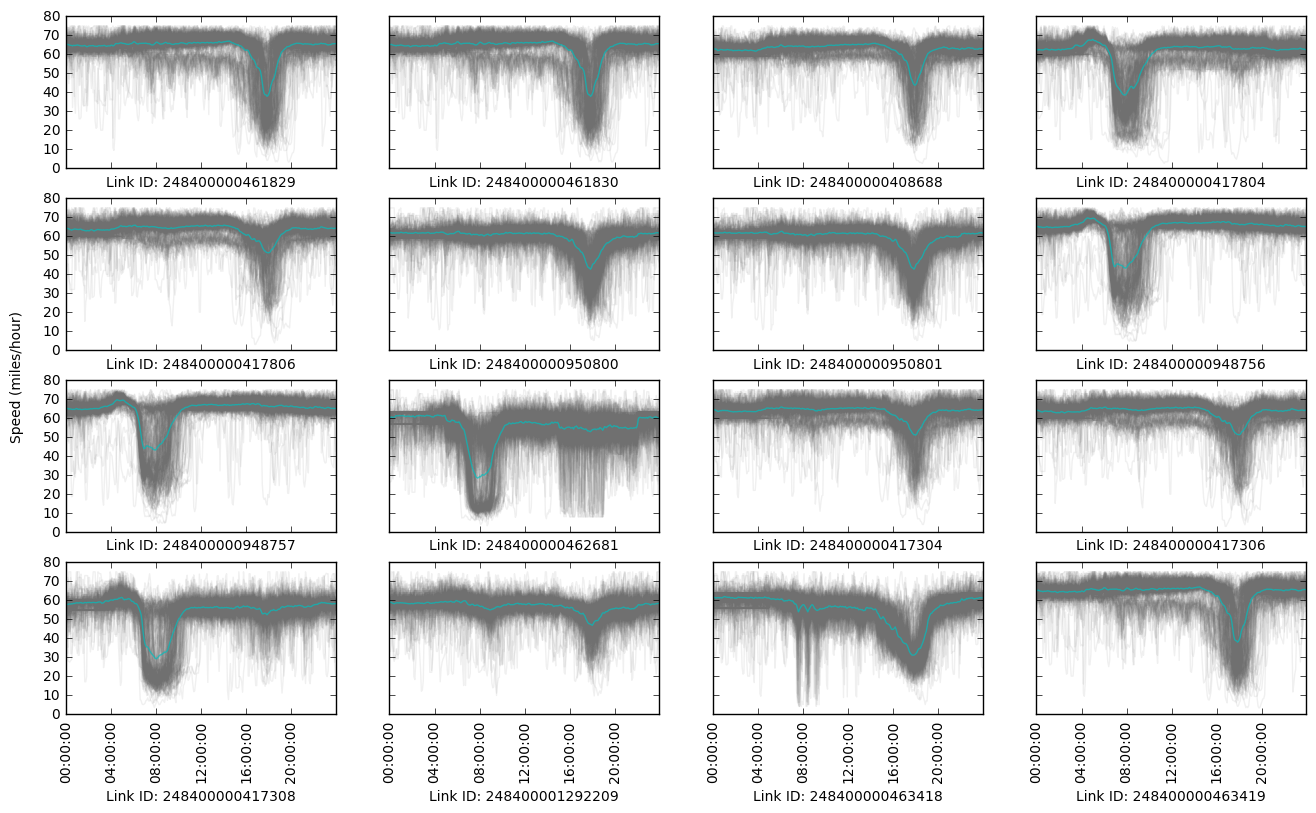}
	\caption{Traffic speed for randomly selected $16$ sensors}
	\label{fig:spd}
\end{figure}

Traffic speed data were obtained from National Performance Management Research Data Set (NPMRDS). The traffic speed data are provided at the geographic level of Traffic Message Channel (TMC), one of the geo-reference protocols. NPMRDS data contain traffic speed observations for $43$ TMCs in every $5$ minutes from $2014$ to $2016$. On each day, there are 288 time intervals, and thus the traffic speed data for each day is a vector in $\mathbb{R}^{288}$. We geocode the TMCs to the network and compute the time-dependent travel time for each road segment. There exist several TMCs attached to the same road segment, we take the average of the traffic speed over those TMCs for that road segment. We visualize the day-to-day traffic speed data for $16$ randomly selected TMCs, as well as the mean time-of-day speed, plot in Figure~\ref{fig:spd}. Each grey time-of-day trace represents traffic speed over one day, and the blue line represents the average traffic speed over three years. Similar pattern as in Figure~\ref{fig:count} can be observed in Figure~\ref{fig:spd}. Similar to counts data, traffic speeds show clearly patterns where speed drops during morning peaks or afternoon peaks, but day-to-day variations are quite large.

There are less than $1\%$ data missing in the speed data. We use linear interpolation across different time intervals on one day and several neighboring days to impute data. For example, if the traffic speed at 10:00 is missing,then  we take the average of traffic speed at 9:55 and 10:05 to impute the traffic speed at 10:00. If data for day 2 are missing, we take the average of traffic data for day 1 and day 3 as the imputed value. Note the former method is always preferred. Only when there are data missing in a large chunk of time intervals, the latter method will be used. 

\subsection{Clustering and route choice analysis}

After processing the data, we use t-SNE to project the dimension of both traffic counts and traffic speed data to a lower dimensional feature space. Then a clustering method is adopted on this feature space to obtain traffic patterns.

\subsubsection{Dimension reduction}

We project both traffic data and speed data to a two-dimensional space so that we can visualize the data easily. TSNE package
in scikit-learn is used to conduct t-SNE algorithm. The parameters for t-SNE are set as follows:
\begin{itemize}
	\item {\em Count data:} perplexity $60$, early exaggeration $12$, learning rate $200$
	\item {\em Speed data:} perplexity $20$, early exaggeration $2$, learning rate $80$
\end{itemize}

\begin{figure}[h]
	\centering
	\begin{subfigure}[b]{0.485\textwidth}
		\includegraphics[width=\textwidth]{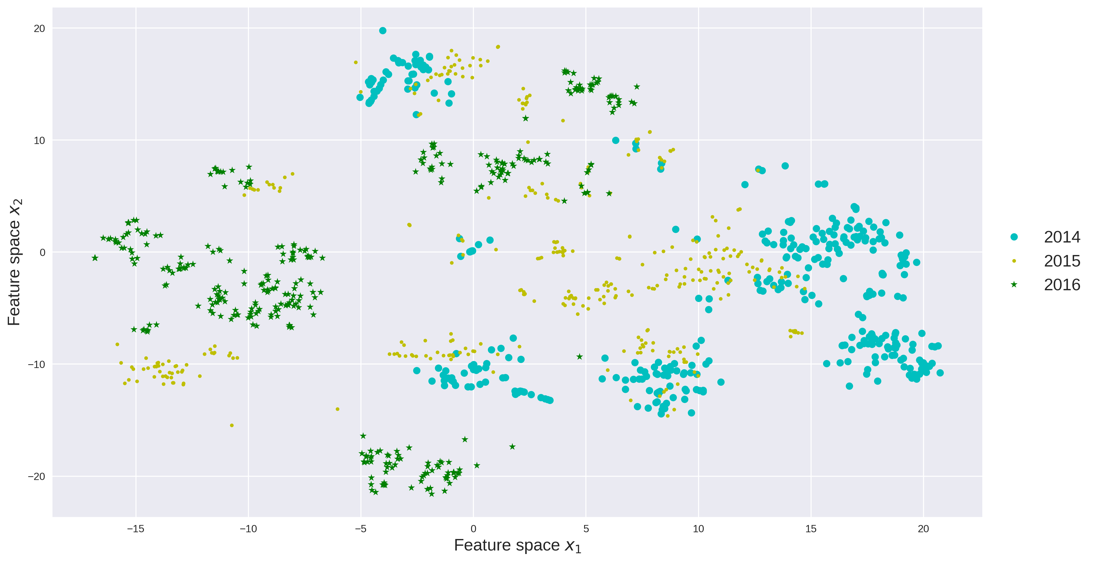}
		\caption{\footnotesize{Yearly pattern of count data}}
		\label{fig:year_count}
	\end{subfigure}
	\begin{subfigure}[b]{0.485\textwidth}
		\includegraphics[width=\textwidth]{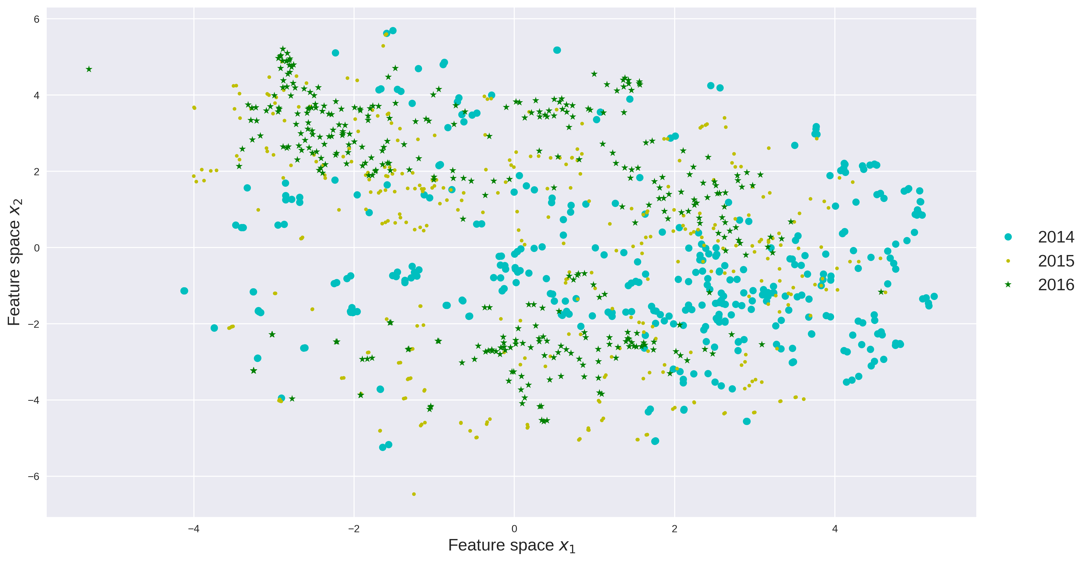}
		\caption{\footnotesize{Yearly pattern of speed data}}
		\label{fig:year_spd}
	\end{subfigure}
	
	\begin{subfigure}[b]{0.485\textwidth}
		\includegraphics[width=\textwidth]{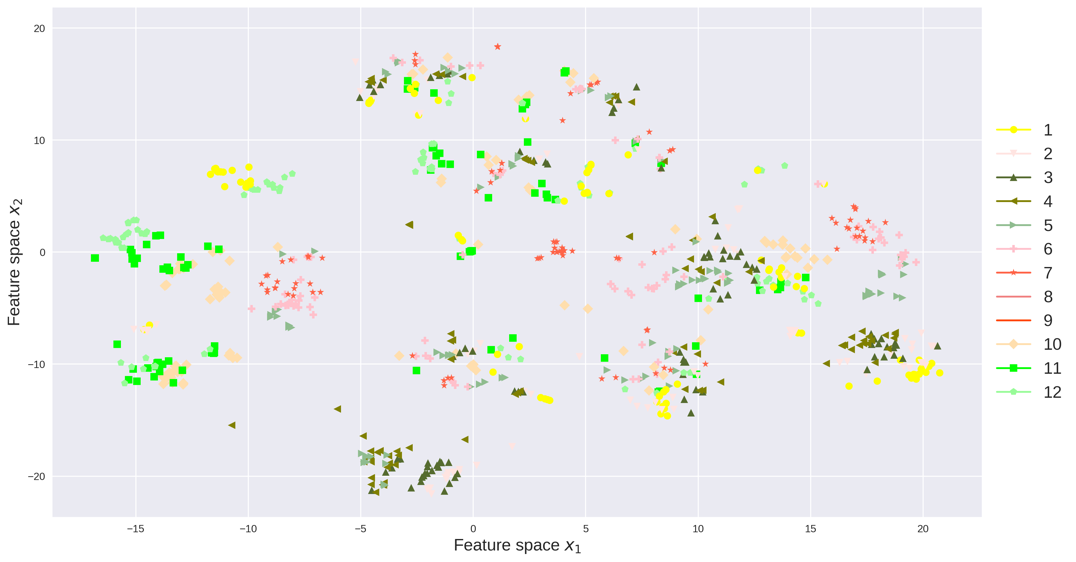}
		\caption{\footnotesize{Monthly pattern of count data}}
		\label{fig:month_count}
	\end{subfigure}
	\begin{subfigure}[b]{0.485\textwidth}
		\includegraphics[width=\textwidth]{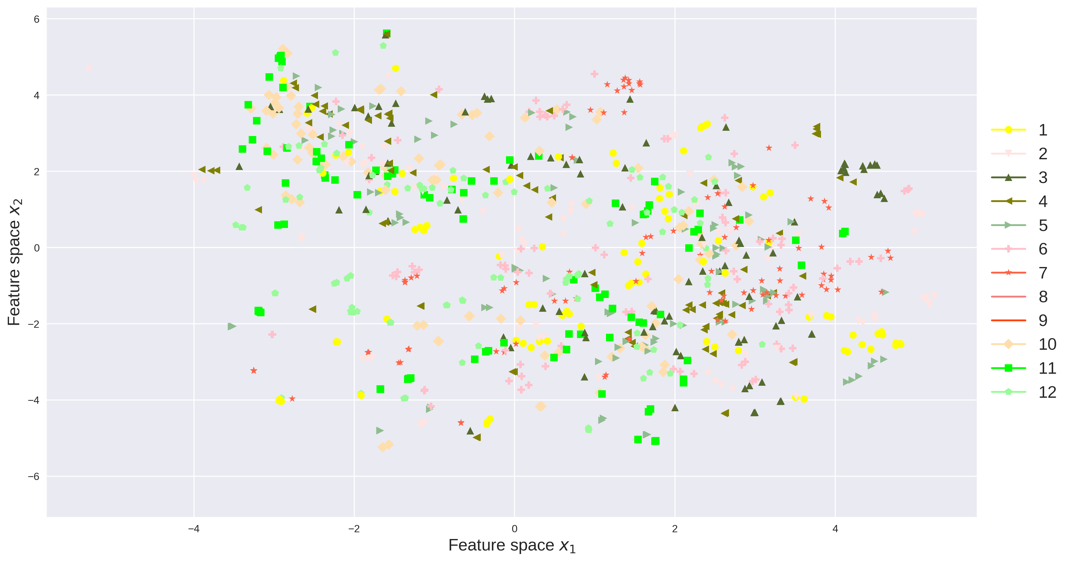}
		\caption{\footnotesize{Monthly pattern of speed data}}
		\label{fig:month_spd}
	\end{subfigure}

	\begin{subfigure}[b]{0.485\textwidth}
	\includegraphics[width=\textwidth]{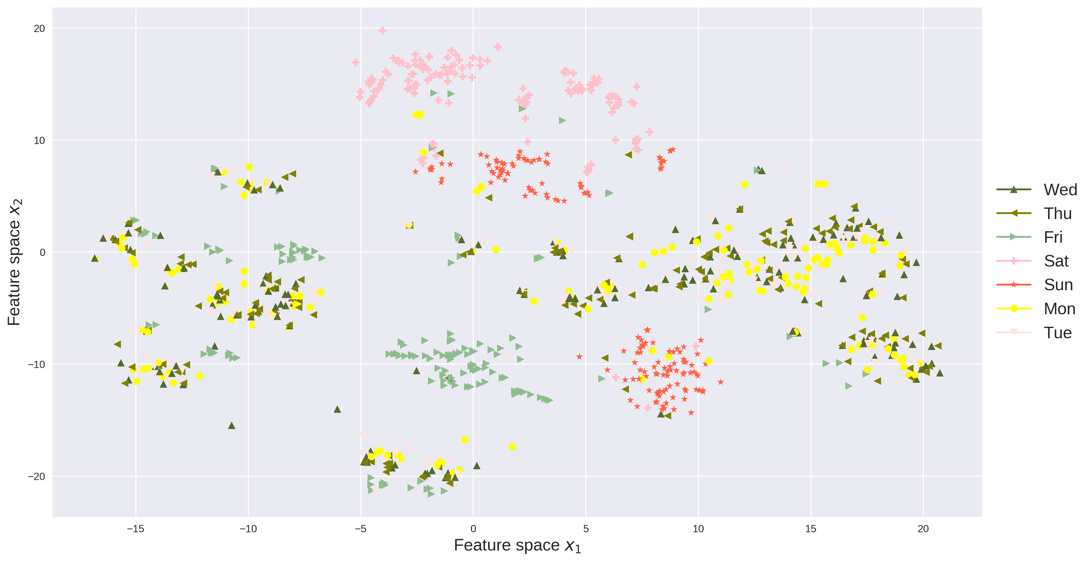}
	\caption{\footnotesize{Weekday pattern of count data}}
	\label{fig:daily_count}
	\end{subfigure}
	\begin{subfigure}[b]{0.485\textwidth}
		\includegraphics[width=\textwidth]{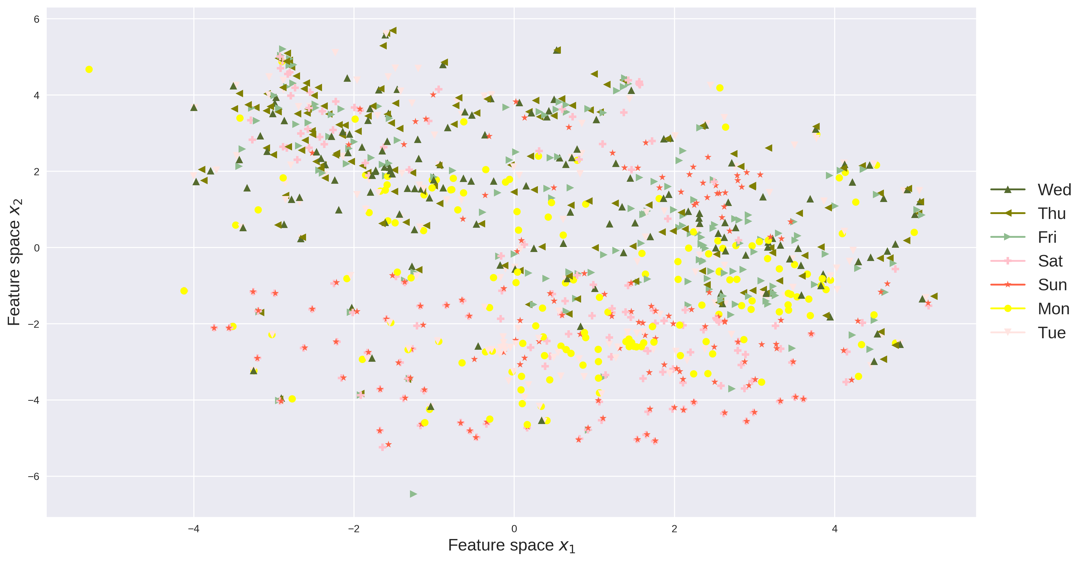}
		\caption{\footnotesize{Weekday pattern of speed data}}
		\label{fig:daily_spd}
	\end{subfigure}

	\caption{Patterns on t-SNE feature space for count and speed data}
	\label{fig:tnse_results}
\end{figure}

The perplexity, early exaggeration and learning rate are parameters in the t-SNE algorithm. These parameters are data dependent and can be tuned through cross validation.  We visualize the count data and speed data in the feature space, respectively. Each point represents traffic data for one day, x-axis and y-axis represent the coordinates of the feature space. The absolute coordinates of each data point does not matter, while the relative positions of these data points matter. The relative positions of the data points indicate whether the data points are similar to each other and how the data points are clustered. We also colored each data point with respect to its year, month and weekday as in Figure~\ref{fig:tnse_results}.

Feature space, like the principle component in PCA, is the base of the low-dimensional space extracted by t-SNE. As can be seen, the count data are more separable as the variance of count data is greater than the variance of speed data. The feature space reflects the yearly, monthly and daily pattern of traffic data. For example in Figure~\ref{fig:year_count} and Figure~\ref{fig:year_spd}, traffic data in 2014 and 2016 are each grouped and far away between each other. Traffic data in 2015 lie in between groups of 2014 and 2016. In Figure~\ref{fig:month_count}, traffic flow in each month is grouped into several clusters, meaning traffic counts data has clearly monthly patterns. While in Figure~\ref{fig:month_spd}, the speed data does not have very clear monthly patterns. Figure~\ref{fig:daily_count} and Figure~\ref{fig:daily_spd} indicate both count data and speed data have strong weekly patterns, as Saturday/Sunday are clustered together and Wednesday/Thursday are clustered together.

We also apply the PCA, Latent Dirichlet Allocation (LDA) and kernel PCA with degree $3$ polynomial kernel to the same count data and speed data, and the weekly/monthly/yearly patterns are not clear from those results. The figures similar to Figure~\ref{fig:tnse_results} can be found in the supplementary materials. The t-SNE tends to divide the data points into small groups, while other methods usually generate a cluttered visualization. To better cluster the data points, we use the results by t-SNE for the rest of the experiments.

\subsubsection{Clustering}

\begin{figure}[h]
	\centering
	\begin{subfigure}[b]{0.485\textwidth}
		\includegraphics[width=\textwidth]{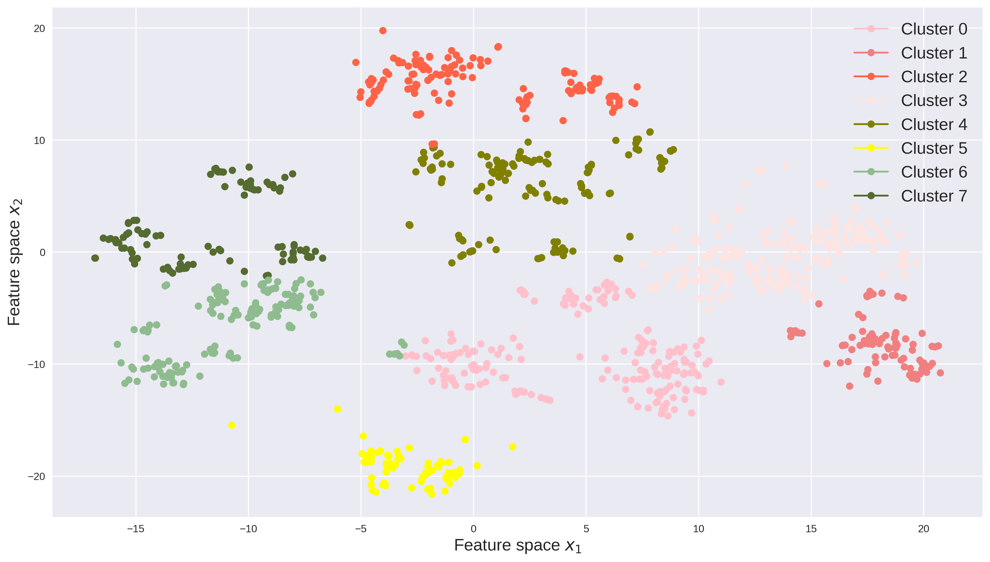}
		\caption{\footnotesize{Count data}}
	\end{subfigure}
	\begin{subfigure}[b]{0.485\textwidth}
		\includegraphics[width=\textwidth]{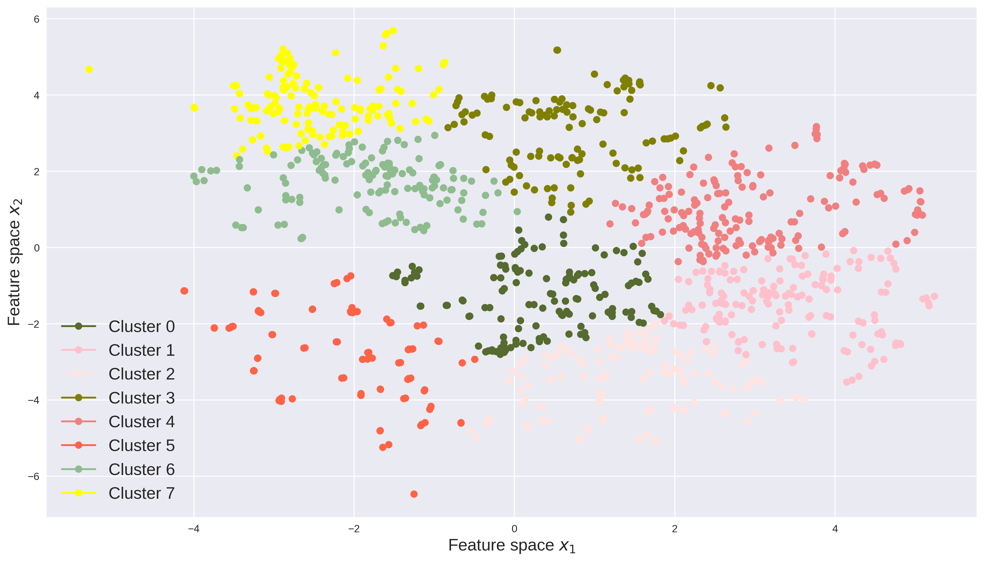}
		\caption{\footnotesize{Speed data}}
	\end{subfigure}
	
	\caption{\footnotesize{Clustering results for count and speed data}}
	\label{fig:cluster_results}
\end{figure}

After dimension reduction, we use k-means to cluster the data points on the feature space. We choose the number of clusters $k = 8$ for both count and speed data, k-means method converges very quickly and the results are shown in Figure~\ref{fig:cluster_results}.

\begin{figure}[h]
	\centering
	\includegraphics[scale = 0.5]{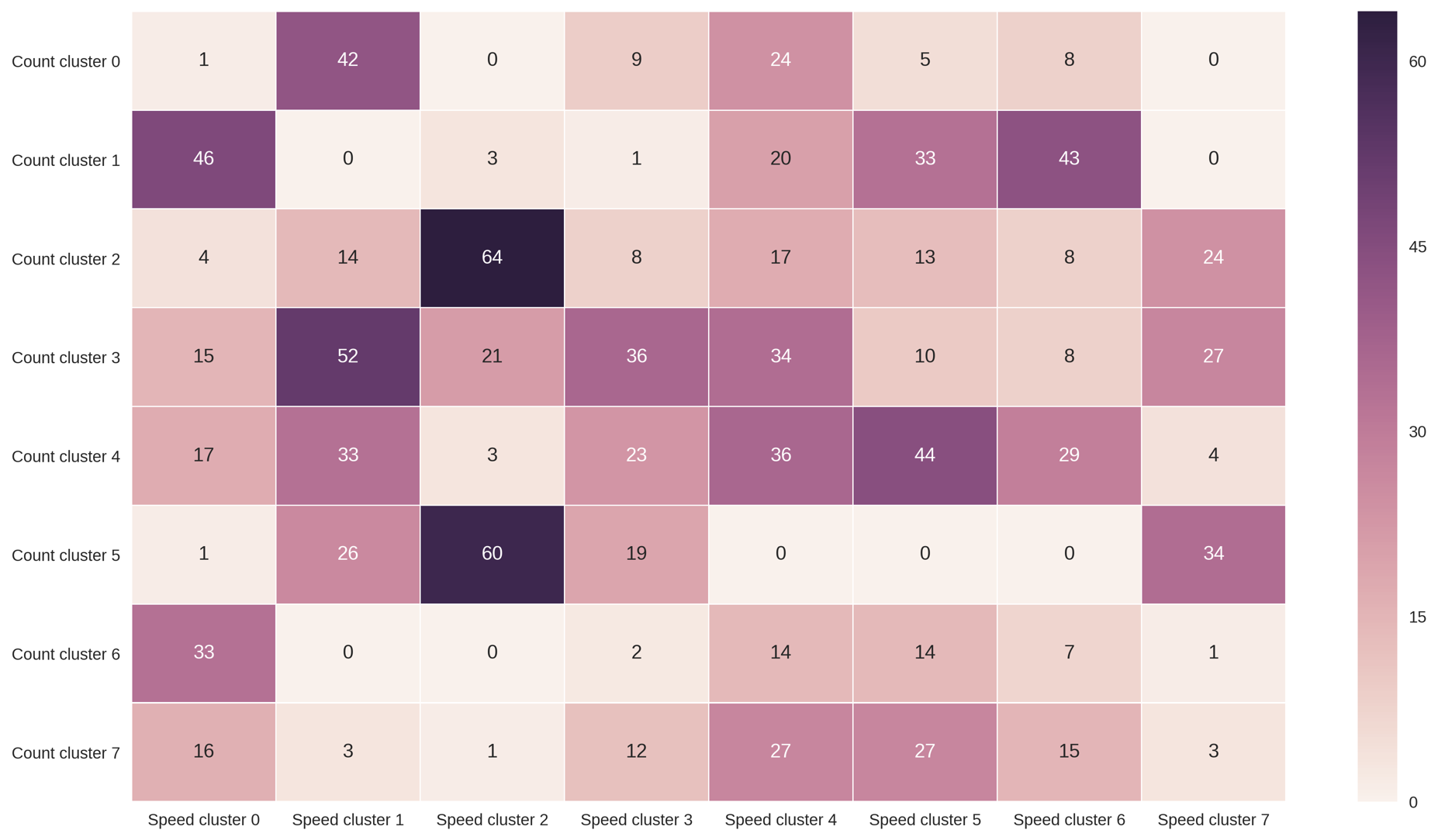}
	\caption{Number of traffic data in each traffic pattern}
	\label{fig:scenatios}
\end{figure}

Travelers can make different route choices based on traffic patterns related to both traffic volumes (traffic counts) or traffic congestion (traffic speed). We define $8 \times 8 = 64$ different traffic patterns to take into account characteristics of different count and  speed clusters. The number of traffic data in each pattern are presented in Figure~\ref{fig:scenatios}. We drop all the patterns with no data point. There are in all $55$ valid traffic patterns.

The outliers are also picked out during the clustering process. For example only one data point falls in the combination of count cluster $0$ and speed cluster $0$. This data point can be viewed as one outlier that does not share similarity with any other traffic patterns. We compute travelers' route choice portions of this outlier day using its unique traffic conditions.

For patterns with more than one data points (i.e., days), we compute the route choice portions using the average traffic speed of all days within each pattern, as discussed in section \ref{sec:route}. We adopt $\theta = 0.01$ since the magnitude of the travel time is around hundreds of seconds. In this demonstrative case study, $\theta$ is determined without careful calibration, which can be improved in the future research using methods proposed by \citet{lu2015enhanced,yang2001simultaneous}.

\subsection{Dynamic OD estimation}

Having the DAR matrix of each day computed by section~\ref{sec:dar} and route choice portion matrix of each pattern computed by section~\ref{sec:route}, we estimate the dynamic OD demand using the proposed stochastic projected gradient descent method.

\subsubsection{Goodness of fit}

In the stochastic gradient method, the configurations are set as follows:
\begin{itemize}
	\item {\em number of epochs}: $300$
	\item {\em batch size}: $8192$
	\item {\em step size}: $5$
	\item {\em use GPU}: True
\end{itemize}

\begin{figure}[h]
	\centering
	\includegraphics[scale = 0.8]{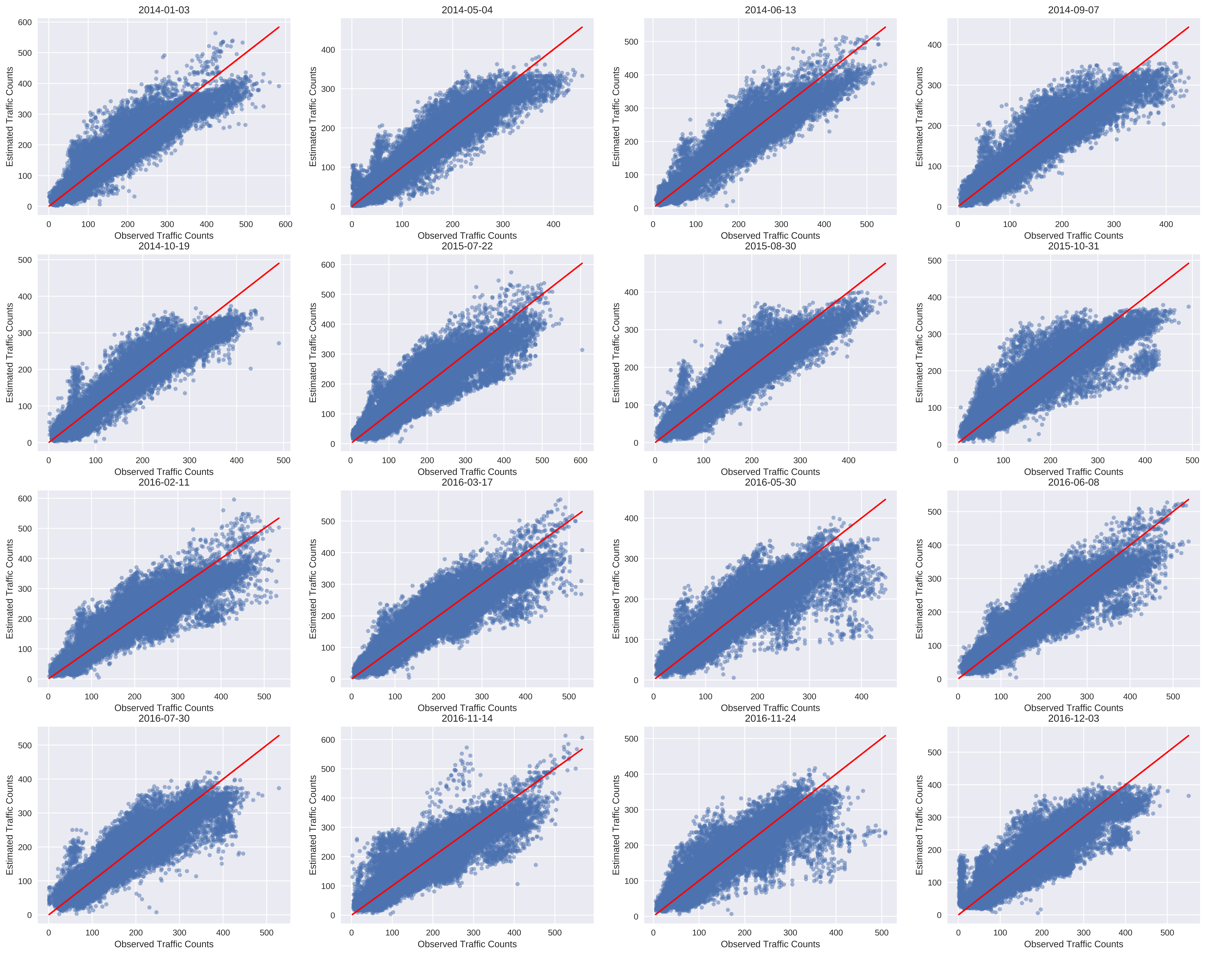}
	\caption{Observed v.s. estimated traffic counts in $16$ randomly selected days}
	\label{fig:fit}
\end{figure}

The entire estimation process for three years takes around $20$ hours, with an average of $1$ minute for each day. We randomly selected $16$ days to visualize the observed traffic counts and estimated traffic counts in Figure~\ref{fig:fit}. The average R-square between the observed link flow and estimated link flow is $0.87$ for three years. The estimated OD demands are able to reproduce the traffic counts observations, implying satisfactory results.

\begin{figure}[h]
	\centering
	\begin{subfigure}[b]{0.485\textwidth}
		\includegraphics[width=\textwidth]{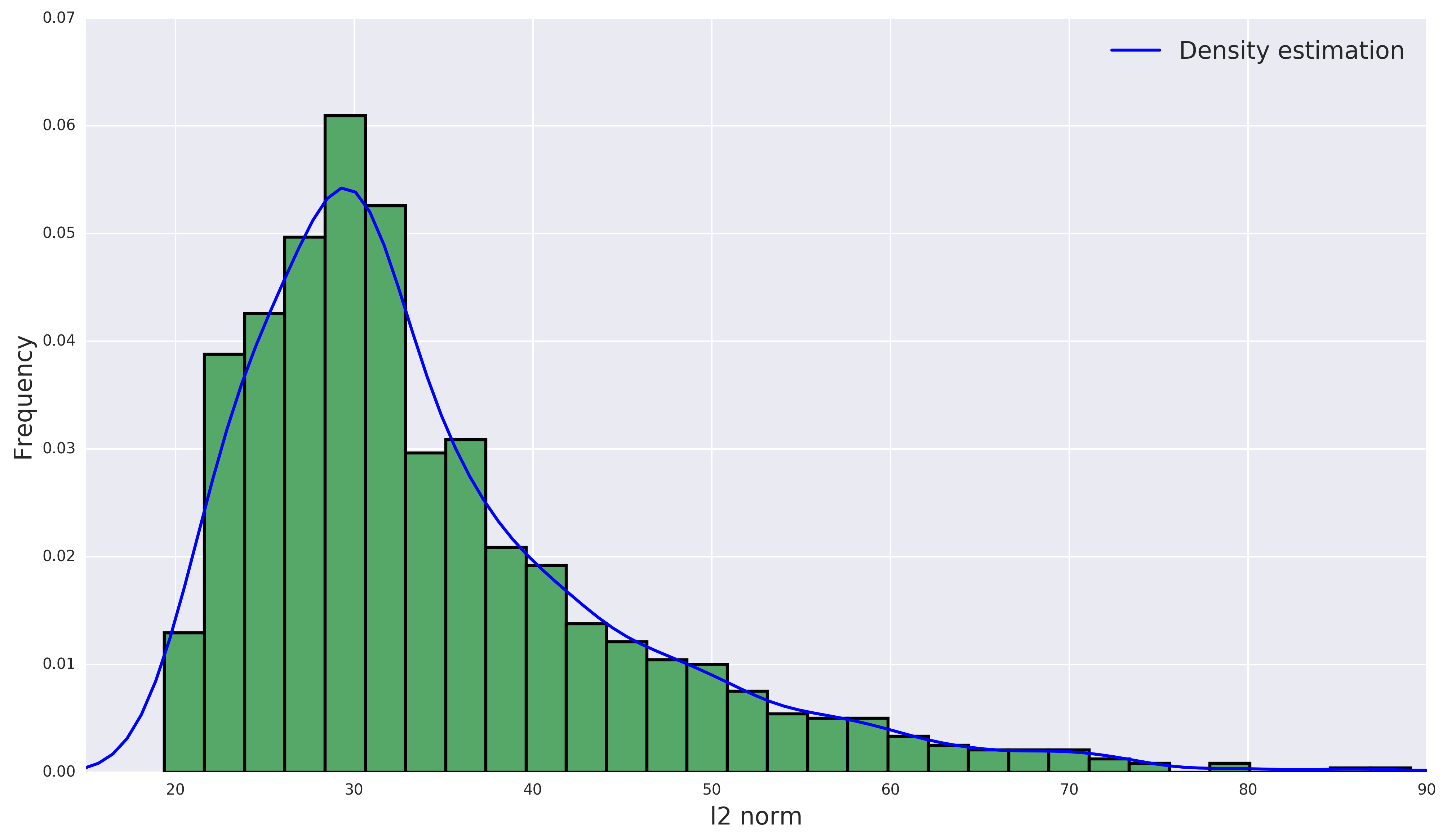}
		\caption{\footnotesize{$\ell^2$ norm distance between the DAR matrix and average DAR matrix over three years}}
		\label{fig:v1}
	\end{subfigure}
	\begin{subfigure}[b]{0.485\textwidth}
		\includegraphics[width=\textwidth]{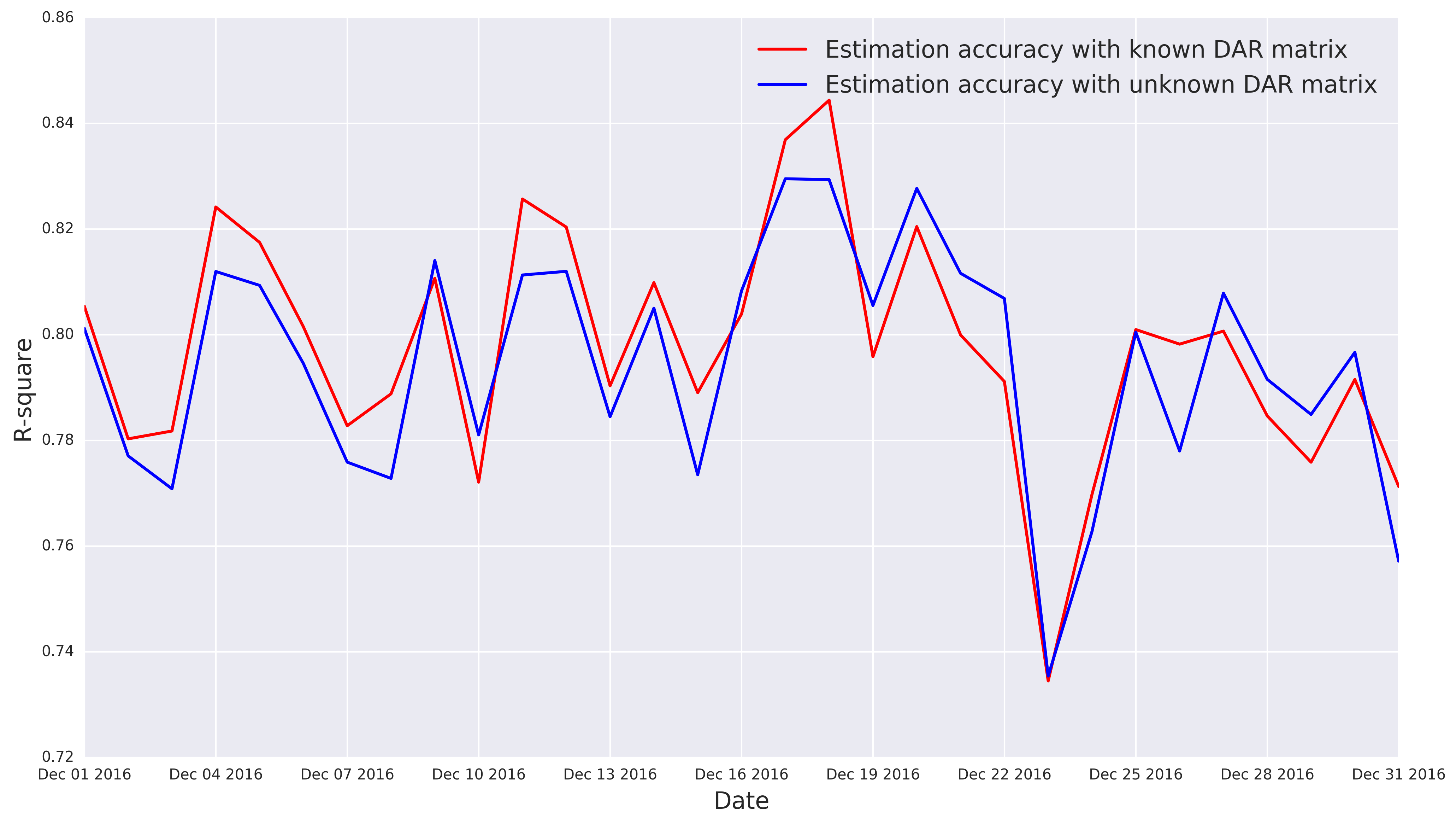}
		\caption{\footnotesize{$R^2$ between the observed link flow and estimated link flow across December, 2018}}
		\label{fig:v2}
	\end{subfigure}
	
	\caption{\footnotesize{Empirical test on the DAR matrix and OD estimation results}}
	\label{fig:v12}
\end{figure}

The true OD demand is difficult to obtain in real-world networks, so the comparison between the estimated OD demand and true OD demand is infeasible in the case study. To further validate the estimation results, we propose a novel interpretation of DODE formulation as follows: we view the observed link flow as the ``data", the DAR matrix as the ``model" and estimated OD as ``target" in the DODE formulation. The terms ``data", ``model" and ``target" are used to assimilate a typical machine/statistical learning task. Under this setting, the DODE formulation can be described as follows: given an observed ``data", we train the ``model" with the speed data and then compute the ``target" by inputting the ``data" to the ``model". We first examine the stability of the ``model". We compute the average DAR matrix across three years and plot the histogram of $\ell^2$ distance between the DAR matrix on each day and the average DAR matrix in Figure~\ref{fig:v1}. One can clearly see the distribution of $\ell^2$ distance is unimodal, which implies the daily perturbation of traffic conditions has a bounded impact to the DAR matrix, thus the OD estimation results are robust to the observation errors and inaccurate DAR matrix.  We also adopt a modified cross-validation approach as follows: we assume the DAR matrices (``model") in December 2018 are unknown and estimated by the average traffic conditions in the other $35$ months. We compute the $R^2$ between the observed link flow and estimated link flow using the estimated DAR matrix and the true DAR matrix, respectively. The results are presented in Figure~\ref{fig:v2}. The DODE with estimated DAR matrix (average $R^2$ is $0.794$) slightly underperforms the DODE with true DAR matrix (average $R^2$ is $0.797$), as expected. The estimation results are still satisfactory, indicating the robustness of the proposed DODE method.

\subsubsection{Algorithm efficiency}

\begin{figure}[h!]
	\centering
	\includegraphics[scale = 0.5]{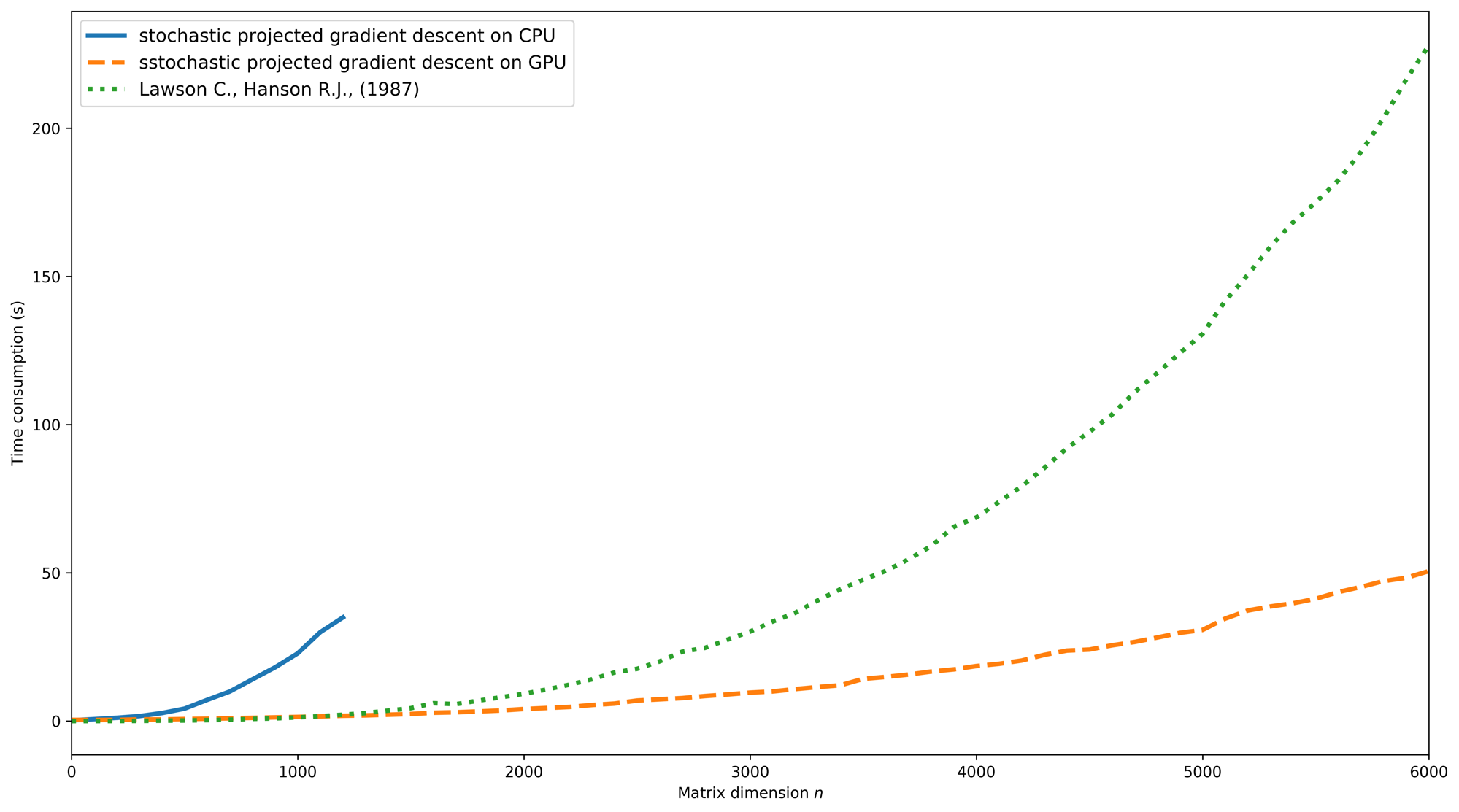}
	\caption{Computation time of three methods with respect to matrix dimensions}
	\label{fig:time}
\end{figure}

We also conduct an experiment to demonstrate the computational efficiency of our proposed algorithm. To compare the CPU based SPGD method, GPU based SPGD and traditional active set based NNLS method \citep{lawson1995solving}, we random generate a matrix $\Beta \in \mathbb{R}^{n \times n}, x \in \left(\mathbb{R}^+\right)^{n}$, we compute $y = \Beta x$ and solve NNLS($\Beta$, $y$) using these three methods. The number of iteration $n$ is set from $100$ to $6000$. As a result, the time consumptions of the three methods are presented in Figure~\ref{fig:time}.

The CPU based SPGD method is very slow so we have to terminate it early. As can be seen, the GPU based SPGD method is significantly the most efficient of all. The gap between standard NNLS method and GPU based gradient project method will increase rapidly as $n$ increases.

In this case study, the dimension of $\Beta$ is $(24768, 23328)$ for the Sacramento regional network. It only takes GPU based SPGD method around $1$ minute to solve it for each day, while the standard active set method will take more than one hour. In this case study, only the GPU based SPGD method can solve the problem of three years in an acceptable amount of time.

\subsection{Aggregated demand over all OD pairs}

With the estimated $5$-minute dynamic OD demand over the three years, we now examine the characteristics of the traffic demand. We start with the aggregated demand over all OD pairs on each day of the three years.

\subsubsection{Weekdays v.s. Weekends}

\begin{figure}[h!]
	\centering
	\begin{subfigure}[b]{0.475\textwidth}
		\includegraphics[width=\textwidth]{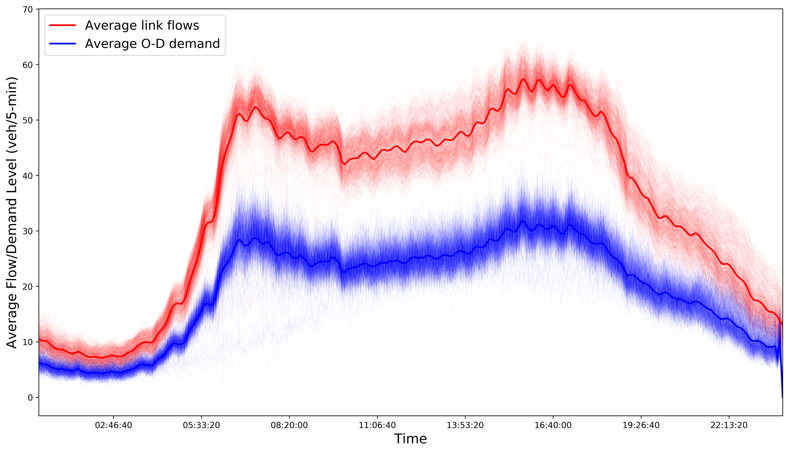}
		\caption{\footnotesize{Weekdays}}
	\end{subfigure}
	\begin{subfigure}[b]{0.475\textwidth}
		\includegraphics[width=\textwidth]{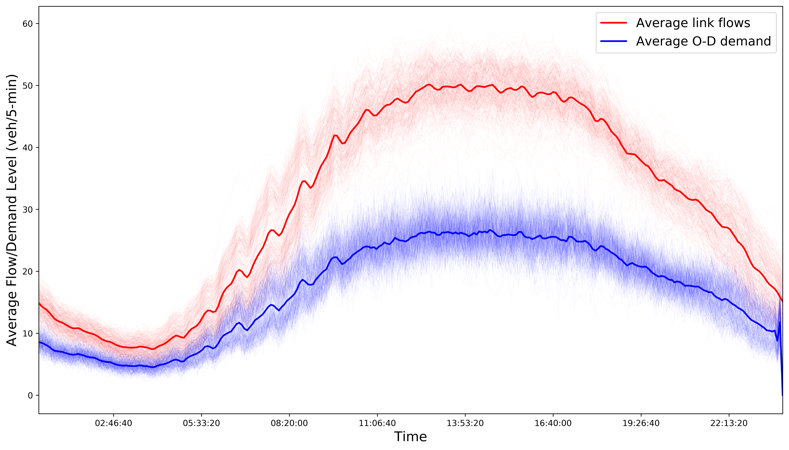}
		\caption{\footnotesize{Weekends}}
	\end{subfigure}
	\caption{\footnotesize{Aggregated OD demand and counts by time of day, on weekdays and weekends (solid lines are the average of aggregated OD demand and counts taken over all weekdays and weekends, respectively)}}
	\label{fig:od_results_week}
\end{figure}

We first look at the differences in aggregated OD demands between weekdays and weekends. For each day, we compute the aggregated OD demand over all OD pairs at each 5-min time interval, and the aggregated traffic counts over all counting locations. Then daily average is computed over the three years. We plot time-of-day aggregated OD and counts for each day (in transparent colors), along with the daily average (in solid colors), in Figure~\ref{fig:od_results_week}. Generally, dynamic OD demand patterns on weekdays and weekends are quite different, as expected. There are two clear spikes on weekdays corresponding to morning and afternoon peaks, respectively. There is only one spike on weekends, and the OD demand on weekends are fairly stable from 11:00am to 17:00pm.

The results show that the aggregated OD demand and aggregated counts have similar time-of-day profiles, but in different scales. Total counts, as commonly used to approximate total demand level in practice, can substantially overestimate the demand level, since they tend to double count the same vehicles that pass through several counting locations. Though both generally follow similar time-of-day profiles, OD demand seems to have spikes and declines slightly earlier than what the total counts read. This indicates that spillover of congestion queues is not too long on both highway corridors, possibly only locally or in the vicinity of a bottleneck.

\subsubsection{Monthly and seasonal effects on OD demand}

\begin{figure}[h]
	\centering
	\includegraphics[scale = 0.8]{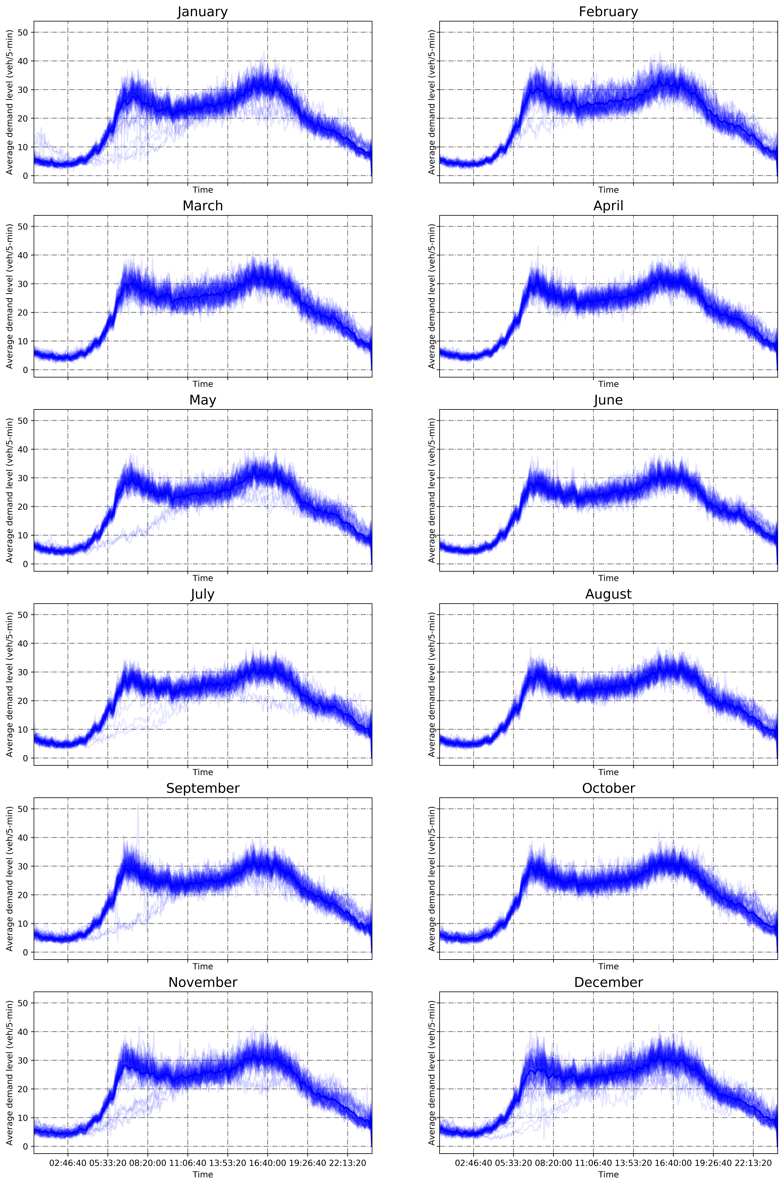}
	\caption{Aggregated OD demand and counts, averaged over all working days in each month}
	\label{fig:od_results_month}
\end{figure}

\begin{figure}[h!]
	\centering
	\includegraphics[scale = 0.4]{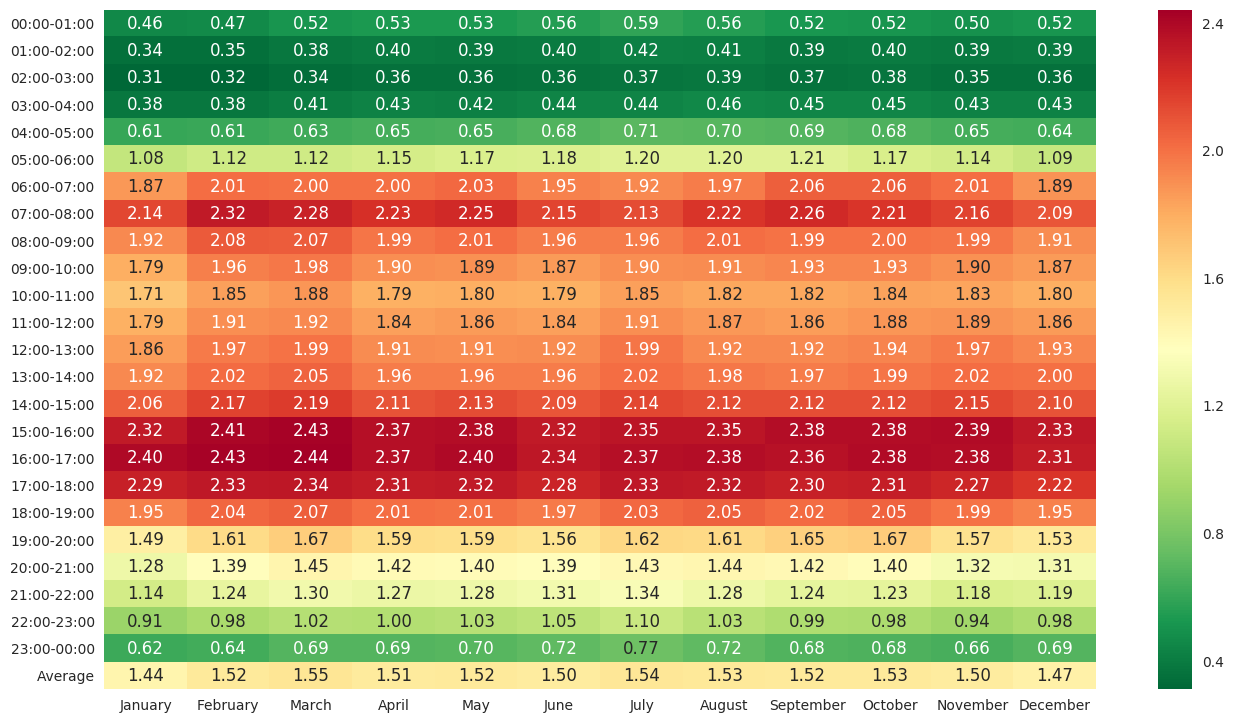}
	\caption{Aggregated OD demand by hour, averaged over all working days in each month ( $\times 10^3$ vehs)}
	\label{fig:abs_month}
\end{figure}

\begin{figure}[h!]
	\centering
	\includegraphics[scale = 0.4]{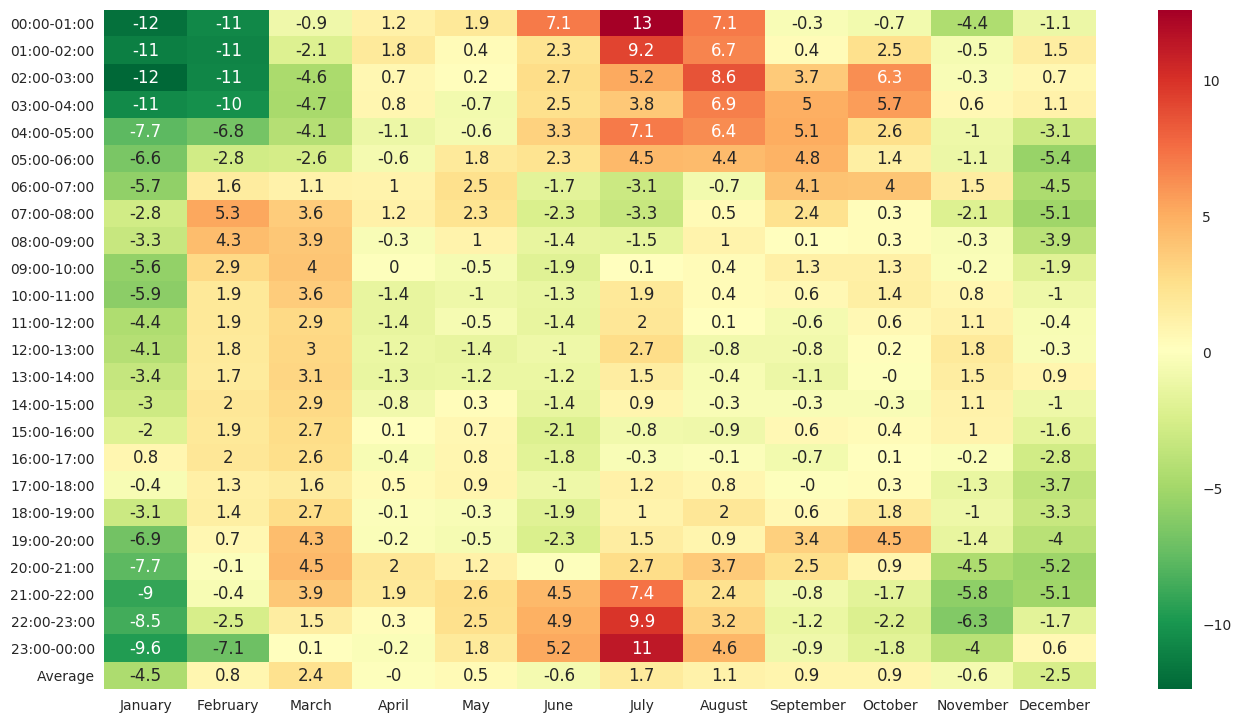}
	\caption{Percentage change in aggregated OD demand by hour by month, comparing to the daily average of aggregated demand taken over all working days of all months ($\%$)}
	\label{fig:per_month}
\end{figure}

For all working days (excluding any holidays on weekdays) in each month, we plot the daily aggregated OD demand over all OD pairs, total counts over all locations, along with their respective daily average for each month, in Figure~\ref{fig:od_results_month}. The general time-of-day profiles are similar across different months. However, the day-to-day variation of OD demand in November, December and January are greater than other months, which may be largely attributed to the travel demands affected by holiday or winter seasons. We also compute the aggregated OD demand by hour, averaged over all working days in each month, in Figure~\ref{fig:abs_month}, as well as the percentage change in aggregated OD demand by hour in Figure~\ref{fig:per_month} where the base is set as the average of aggregated OD demand taken over all months.

OD demands during the morning peaks in June - August and December - January are slightly lower than other months, resulting less congestion during morning peaks. Among those, morning peak demand in July drops the most considerably compared to other months. On the other hand, summer time (from May to September) shows higher demand during off-peak hours, especially July and August. Overall, the total travel demand in December and January are the lowest throughout the years. Those monthly and seasonal demand change may be related to the summer/winter breaks of schools, and effects of summer/winter weather. These phenomena are consistent with our perception, and can be demonstrated and validated by three years' data, which cannot be discovered by examining speed/counts data directly.

\subsubsection{Northbound v.s. Southbound}

We plot the aggregated OD demand by weekdays and weekends, and over all northbound and southbound OD pairs, respectively, in Figure~\ref{fig:od_results_month}.

\begin{figure}[h]
	\centering
	\begin{subfigure}[b]{0.485\textwidth}
		\includegraphics[width=\textwidth]{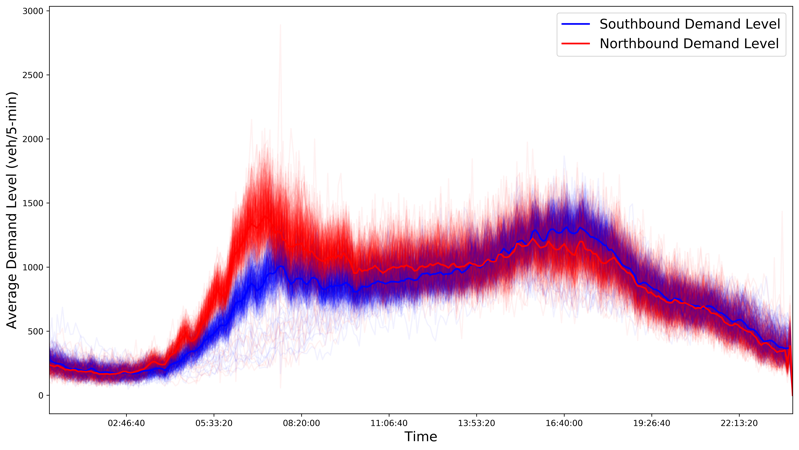}
		\caption{\footnotesize{Weekdays}}
	\end{subfigure}
	\begin{subfigure}[b]{0.485\textwidth}
		\includegraphics[width=\textwidth]{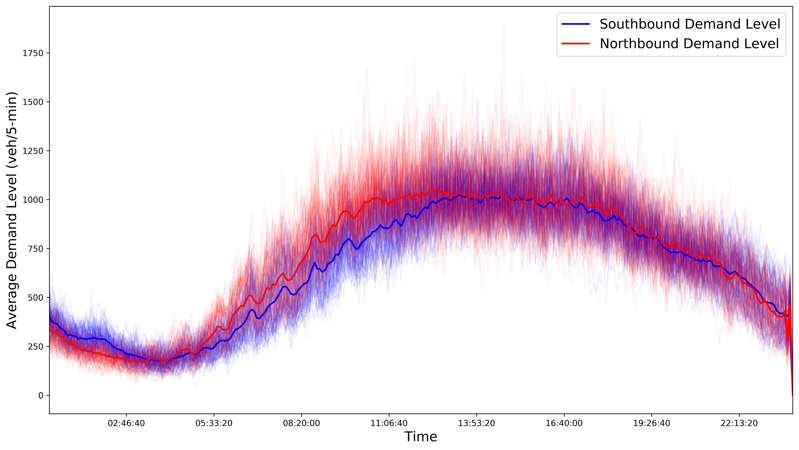}
		\caption{\footnotesize{Weekends}}
	\end{subfigure}
	\caption{\footnotesize{Aggregated OD demand, by northbound and southbound}}
	\label{fig:od_results_ns}
\end{figure}

Northbound demand heads to the Sacramento downtown, and southbound demand heads to the southern region. On weekdays, the northbound OD demand is greater than southbound OD demand during morning peaks, and slightly less during afternoon peaks. Morning commute clearly shows more day-to-day variation than other time periods. One interesting observation is that the discrepancy between northbound/southbound OD demand in afternoon peaks is less than that in morning peaks. Congestion during the day is usually more widely spread than morning commute congestion that mainly applies to northbound only.


On weekends, the OD demand per hour is considerably less than the demand rate during morning commute on weekdays. Northbound sees a higher demand level and earlier weekend peak than southbound. However, during midnight, more demand travels on southbound than northbound, possibly as a result of midnight activities in Sacramento Downtown.


\subsubsection{Holidays v.s. weekdays immediately after holidays}

\begin{figure}[h]
	\centering
	\includegraphics[scale = 1]{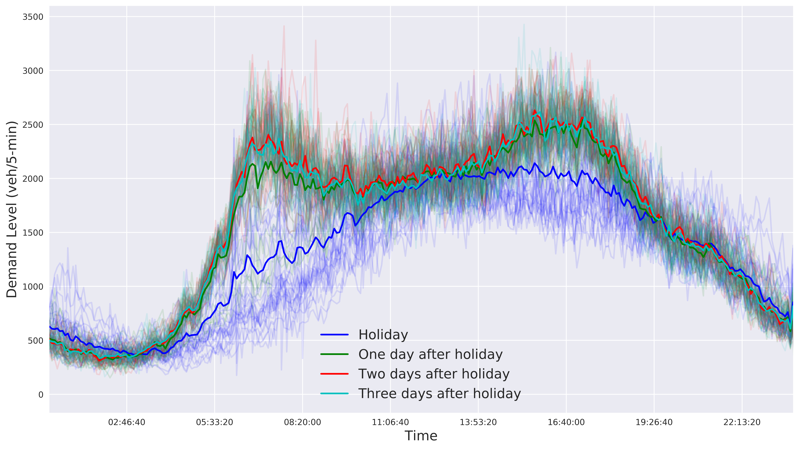}
	\caption{Aggregated OD demand, on holidays and on weekdays immediately after holidays}
	\label{fig:holiday}
\end{figure}

OD demand during holidays appears quite different comparing to the regular weekdays and weekends. Thus, we pick out all the holidays (excluding the weekends), and those working days immediately after holidays to visualize their respective demand patterns. For example, September 5 2016 is a Labor day on Monday, then September 6 2016 is one weekday immediately after the holiday. We compute the aggregated OD demand for the two types, and present the results in Figure~\ref{fig:holiday}.

As can be seen from Figure~\ref{fig:holiday}, holiday traffic patterns are closer to the weekend patterns then to the weekday patterns, with one big spike during the day. However, a small morning peak can exist for some holidays, possibly attributed to different nature of daytime activities from a regular weekend. Another interesting finding for the holiday OD demand pattern is that the midnight OD demand can be as high as $1,250$, almost half of the aggregated demand during morning peaks.

Though a morning commute peak resumes after holidays, we see that the peak on the weekday immediately after holidays is considerably lower than that of a regular weekday. OD demand patterns become normal from the second weekday after the holidays.

\subsection{Disaggregated demand}
Now we examine 24/7 OD demand of each OD pair over the $3$ years.

\subsubsection{Northbound v.s. Southbound}

\begin{figure}[h]
	\centering
	\includegraphics[scale = 1]{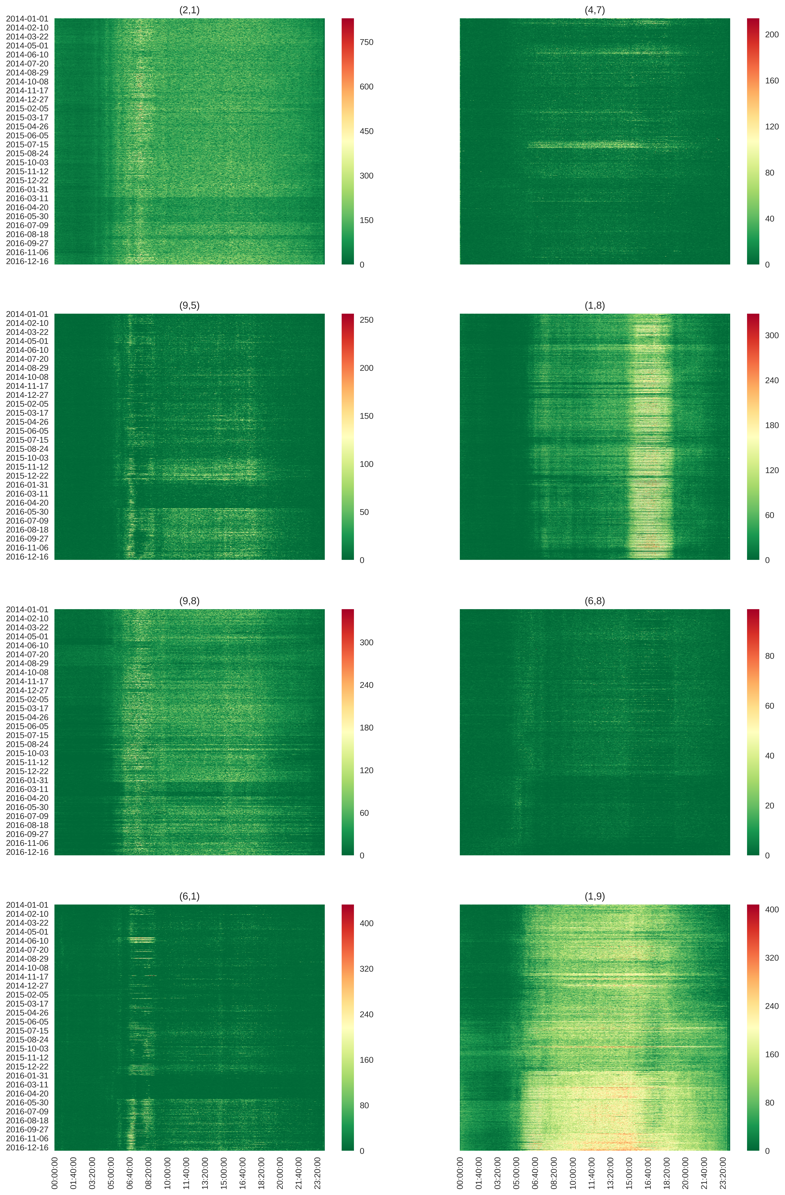}
	\caption{Time-of-day OD demand profile for randomly selected northbound/southbound OD pairs}
	\label{fig:detail_ns}
\end{figure}

We draw a figure with $(n \times m)$ pixels, $n$ is the number of days and $m$ is the number of time intervals on each day. We set y axis to be the dates from $2014$ to $2016$, and x axis to be the time of day from $00:00$ to $23:59$. Each pixel is color coded to indicate the OD demand level. This figure demonstrates the daily time-of-day demand change over the years for each OD pair in high granularity. We randomly selected $4$ northbound and $4$ southbound OD pairs, and plot them in Figure~\ref{fig:detail_ns}. OD demand between the zone $(1, 9)$ has increased substantially especially during the year of $2016$, resulting an increased demand level throughout the entire 24 hours. Also for OD pair $(6,1)$, there are clearly $3$ spikes during morning commute, and demand for morning commute increases considerably in $2016$. However, other OD pairs plot in Figure~\ref{fig:detail_ns} do not necessarily witness demand increase over time.

One can clearly see that there exist some strips with green color, implying temporary effects on travel demand for some OD pairs. For instance, OD demand is significantly reduced during Jan-Apr 2016 between the OD pair $(6,1), (9,5)$. This could be possibly induced by construction projects in the regional networks that have more impacts on those OD pairs than others.

\subsubsection{Mean and variance of dynamic OD demand}

\begin{figure}[h]
	\centering
	\begin{subfigure}[b]{0.485\textwidth}
		\includegraphics[width=\textwidth]{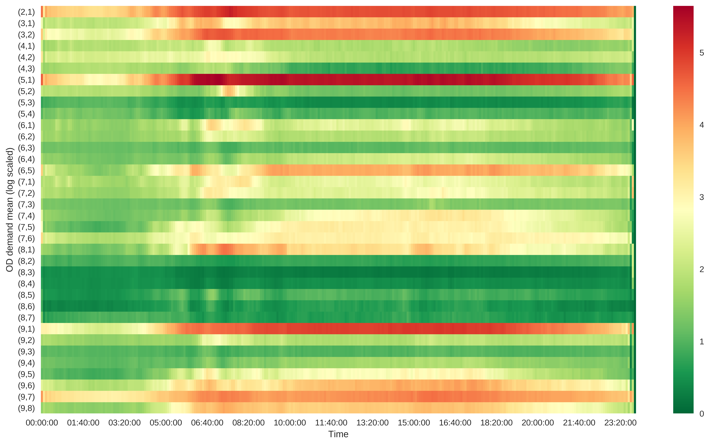}
		\caption{\footnotesize{Northbound OD mean}}
	\end{subfigure}
	\begin{subfigure}[b]{0.485\textwidth}
		\includegraphics[width=\textwidth]{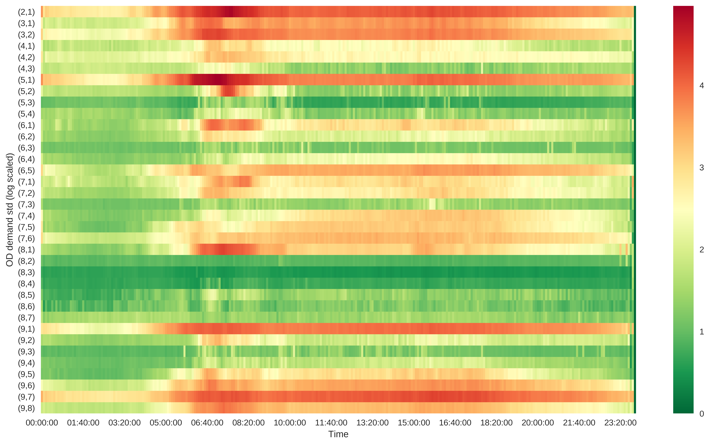}
		\caption{\footnotesize{Northbound OD standard deviation}}
	\end{subfigure}
	\begin{subfigure}[b]{0.485\textwidth}
	\includegraphics[width=\textwidth]{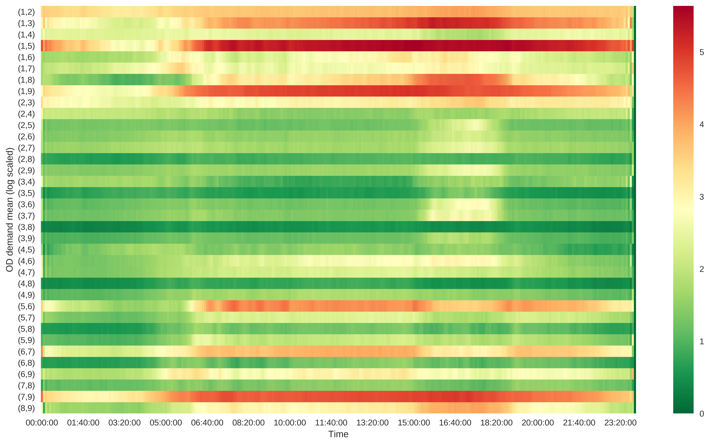}
	\caption{\footnotesize{Southbound OD mean}}
	\end{subfigure}
	\begin{subfigure}[b]{0.485\textwidth}
		\includegraphics[width=\textwidth]{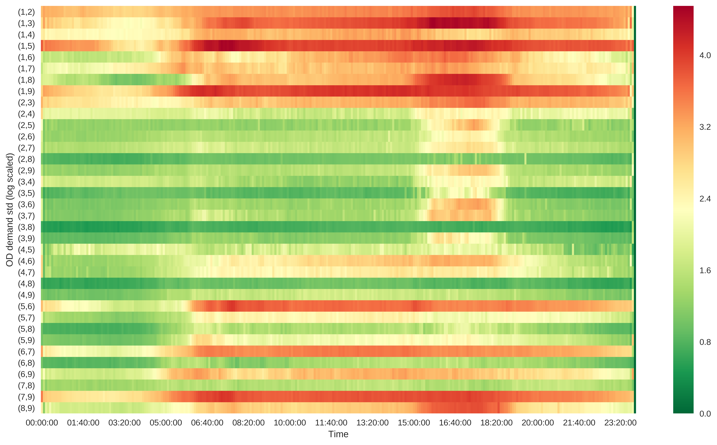}
		\caption{\footnotesize{Southbound OD standard deviation}}
	\end{subfigure}
	\caption{\footnotesize{Mean and variation of OD demand, by OD pair and time of day}}
	\label{fig:od_results}
\end{figure}

We compute the average and standard deviation of each OD pair for each 5-min time interval over $3$ years, and plot them on a heatmap in Figure~\ref{fig:od_results}. We set y-axis to be each OD pair, x-axis to be the time from $00:00$ to $23:59$. Each pixel is color coded to indicate the OD demand level.

As can be seen from Figure~\ref{fig:od_results}, the mean and variance of each OD pair roughly follow similar patterns, and the variance increases with respect to the increase in mean. Origin zones $1, 5, 6,7$ are the most important origins generating demand for southbound direction. Similarly, origin zones $2,5,8,9$ are the important demand origins for northbound direction.

In addition, there exist several OD pairs, such as $(4,1)$, $(1,6)$, with low demand mean and relatively high flow variability. The high variability of the demand among these OD pairs may be caused by accidents or events, so in a way, they may be more vulnerable under non-recurrent traffic conditions.

\begin{figure}[h]
	\centering
	\includegraphics[scale = 1]{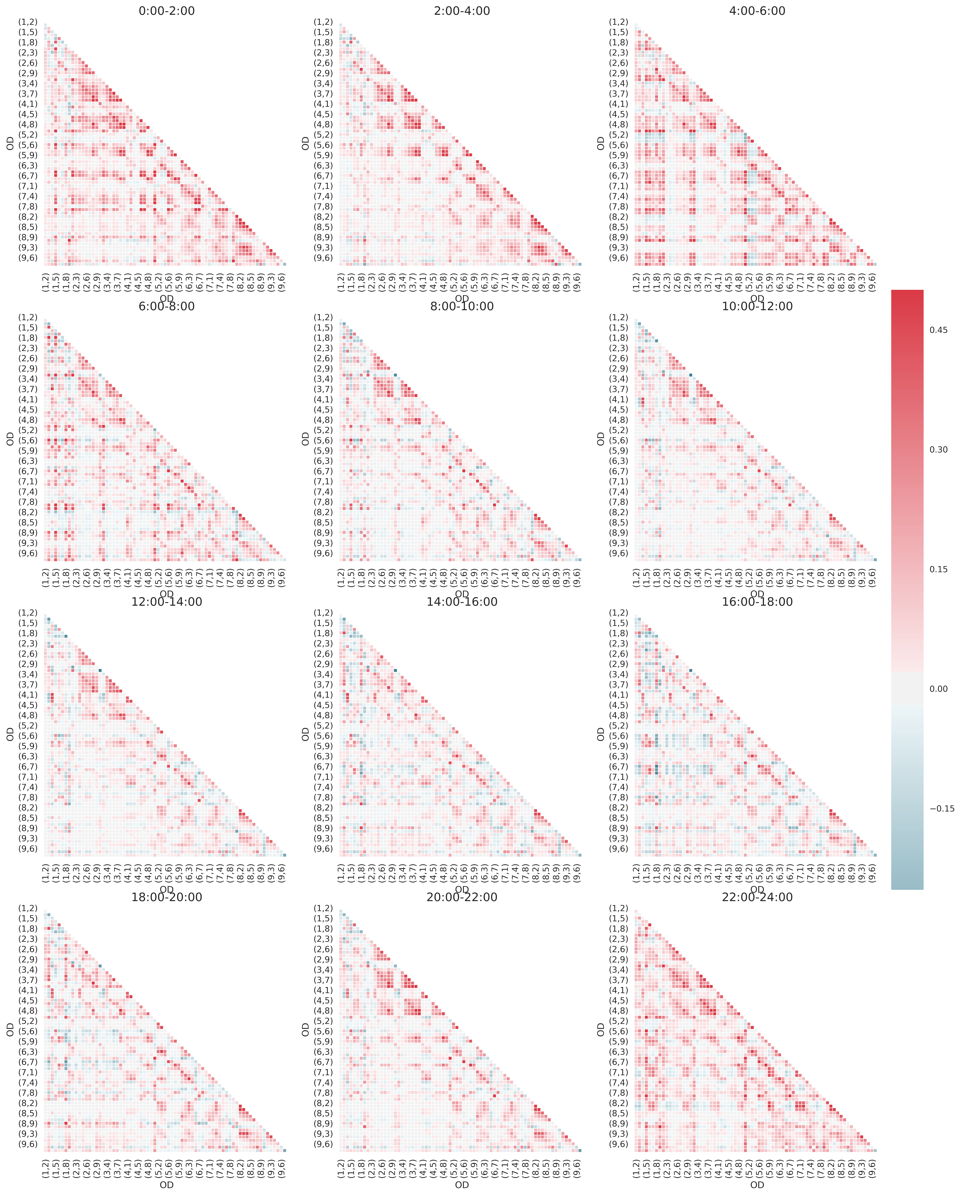}
	\caption{OD demand correlation for different time intervals}
	\label{fig:covariance}
\end{figure}

The correlation between OD pairs is useful when making the transportation planning policies. We compute the Pearson correlation factor between all OD pairs by time of day, and present the results in Figure~\ref{fig:covariance}. The demand among majority of OD pairs is positively correlated. Only a small portion of OD pairs are negatively correlated, which may be worth further investigating the reasons. Generally correlations are higher during peak hours and midnight than those from 10:00 to 16:00.

\subsubsection{Holidays v.s. weekdays immediately after holidays}

\begin{figure}[h]
	\centering
	\includegraphics[scale = 1]{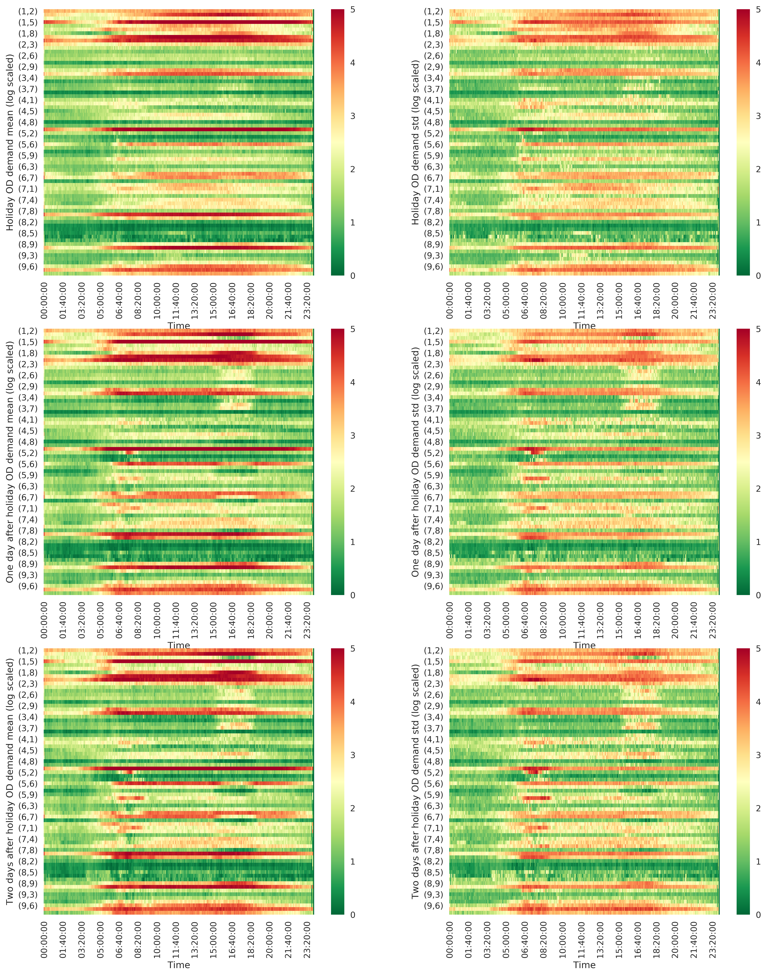}
	\caption{\footnotesize Day-to-day OD demand mean and variance on holidays and weekdays immediately after holidays (left: mean; right: standard deviation; the first row: holidays; the second row: the first weekday after holidays; the third row: the second weekday after holidays)}
	\label{fig:detail_holiday}
\end{figure}

We visualize the day-to-day mean and variance of OD demand for each OD pair on holidays and two weekdays immediately after holidays in Figure~\ref{fig:detail_holiday}. The results are consistent with before, generally demand variance increases with respect to the mean for each OD pair. There is no significant morning or afternoon peak hours for holiday travel demand. Though the total OD demand level on holidays is lower than weekdays, the holiday demand variance is much higher. The first weekday after holidays and the second weekday after holidays follow a similar pattern, while the latter demand is overall higher than the former demand. This again validates our finding for the aggregated OD demand.

\cleardoublepage
\section{Conclusion}
\label{sec:con}

This paper proposes a data-driven framework for estimating multi-year 24/7 dynamic OD demand using high-granular traffic counts and speed data. The proposed framework defines a dynamic assignment ratio (DAR) matrix to encapsulate the traffic flow dynamics and congestion spill-over in the large-scale network. The DAR matrix can be calibrated through high-granular speed data (such as probe vehicle speeds), which alleviates the complexity of non-linear large-scale network simulation for DODE.

The purposed framework adopts t-SNE and k-means methods to reduce the dimensionality of multi-source high-granular data, and cluster those data into typical daily traffic patterns. The t-SNE method projects the multi-source data onto a low dimensional feature space that enables examination of the daily, weekly and monthly patterns of traffic data. The k-means method clusters the projected counts and speed data into traffic patterns. The framework works with any general route choice models that considers day-to-day and within-day travel time and cost. In particular, a Logit-based route choice model is demonstrated to compute the route choice portions under each traffic patterns separately.

The DODE framework can be cast into a standard non-negative least square (NNLS) problem with, however, very high dimensions provided with high-granular data. A novel stochastic projected gradient descent (SPGD) method is purposed to solve for NNLS. The SPGD method can be implemented on GPU, which is able to solve the high dimensional NNLS efficiently compared to the traditional active set method for the NNLS problem. The entire solution framework is implemented in Python and open sourced.

Finally, a case study is conducted on a regional Sacramento network consisting with I-5 and SR-99 corridors, interchanges and ramps. High-granular counts and speed data are used to estimate $5$-minute dynamic OD demands over the three years from 2014 to 2016. The estimation takes around $20$ hours on an inexpensive GPU-based desktop. The estimated dynamic OD demand can fit the large-scale high-granular data fairly well. We also examine daily, monthly, seasonal and yearly changes in OD demand that vary by time of day, by holidays, weekdays and weekends. Those new information regarding travel demand can help city planners and policymakers better understand the characteristics of dynamic OD demands and their evolution/trends in the past few years. The estimated dynamic OD can also be used to compute the variability of day-to-day OD demand, a critical input for network reliability studies \citep{li2018adaptive}.

\section*{Supplementary materials}

The proposed framework is implemented in Python and open-sourced on Github\footnote{\url{https://github.com/Lemma1/DPFE}}. The Github repository also contains the dimension reduction results by PCA, Latent Dirichlet Allocation (LDA) and kernel PCA with degree $3$ polynomial kernel.

\section*{Acknowledgements}
This research is funded in part by National Science Foundation Award CMMI-1751448 and Carnegie Mellon University's Mobility21, a National University Transportation Center for Mobility sponsored by the US Department of Transportation. The contents of this report reflect the views of the authors, who are responsible for the facts and the accuracy of the information presented herein. The U.S. Government assumes no liability for the contents or use thereof.

\cleardoublepage
\bibliography{report}
\cleardoublepage

\end{document}